\documentclass[aps,pre,twocolumn,superscriptaddress,showpacs,longbibliography]{revtex4-1}

\usepackage{amsfonts,amssymb,amsmath,latexsym,epsfig,wasysym} 
\usepackage[sort&compress]{natbib}
\usepackage{color}
\definecolor{bluecolor}{rgb}{0,0.,1.}

\definecolor{redcolor}{rgb}{.7,0.,0.}

\usepackage{bm}
\newcommand{\pp}{\bm{p}}
\newcommand{\qq}{\bm{q}}
\newcommand{\PP}{\bm{P}}

\begin{document}

\title{Similarity of symbol frequency distributions with heavy tails}
\author{Martin Gerlach} 
\affiliation{Max Planck Institute for the Physics of Complex Systems, D-01187 Dresden, Germany}

\author{Francesc Font-Clos} 
\affiliation{Centre de Recerca Matem\`atica,
Edifici C, Campus Bellaterra,
E-08193 Barcelona, Spain.
} 
\affiliation{Departament de Matem\`atiques, 
Facultat de Ci\`encies,
Universitat Aut\`onoma de Barcelona,
E-08193 Barcelona, Spain}

\author{Eduardo G. Altmann} 
\affiliation{Max Planck Institute for the Physics of Complex Systems, D-01187 Dresden, Germany}

\begin{abstract}
Quantifying the similarity between symbolic sequences is a traditional problem in
Information Theory which requires comparing the frequencies of symbols in different sequences.  In numerous modern applications, ranging from DNA over music
to texts, the distribution of symbol frequencies is characterized by heavy-tailed
distributions (e.g., Zipf's law). The large number of low-frequency symbols in these distributions poses
major difficulties to the estimation of the similarity between sequences, e.g., they
hinder an accurate finite-size estimation of entropies.  
Here we show analytically how the systematic (bias) and statistical (fluctuations)
  errors in these estimations depend on the sample size~$N$ and on the exponent~$\gamma$ of the
  heavy-tailed distribution. Our results are valid for the Shannon entropy $(\alpha=1)$,
  its corresponding similarity measures (e.g., the Jensen-Shanon divergence), and also for
  measures based on the generalized entropy of order $\alpha$.  
For small $\alpha$'s, including $\alpha=1$, the errors decay slower than the $1/N$-decay observed in short-tailed distributions. 
For $\alpha$ larger than a critical value $\alpha^* = 1+1/\gamma \leq 2$, the $1/N$-decay is recovered. 
We show the practical significance of our results by quantifying the evolution of
the English language over the last two centuries using a complete $\alpha$-spectrum of measures. We find that frequent words change more slowly than
less frequent words and that $\alpha=2$ provides the most robust measure to quantify language change.

\end{abstract} 

% \pacs{89.65.-s, 89.75.Da, 87.23.Ge, 05.10.-a}
% 05.10.-a        Computational methods in statistical physics and nonlinear dynamics (see also 02.70.-c in mathematical methods in physics)
% 87.23.Ge        Social systems ecology and evolution (Dynamics of Social Systems)
% 89.65.-s        Social and economic systems
% 89.70.-a        Information and communication theory
% 89.75.-k        Complex systems (for complex chemical systems, see 82.40.Qt; for biological complexity, see 87.18.-h)
% 89.75.Da        Systems obeying scaling laws, Scaling phenomena in comlpex systems
% 89.70.Cf	  Entropy in information theory
\maketitle

%%%%%%%%%%%%%%%%%%%%%%%%%%%%%%%%% Introduction %%%%%%%%%%%%%%%%%%%%%%%%%%%%%%%%%
%\showthe\columnwidth
\section{Introduction}\label{sec.intro}

\begin{figure}[bt]
\centering
\includegraphics[width=1.0\columnwidth]{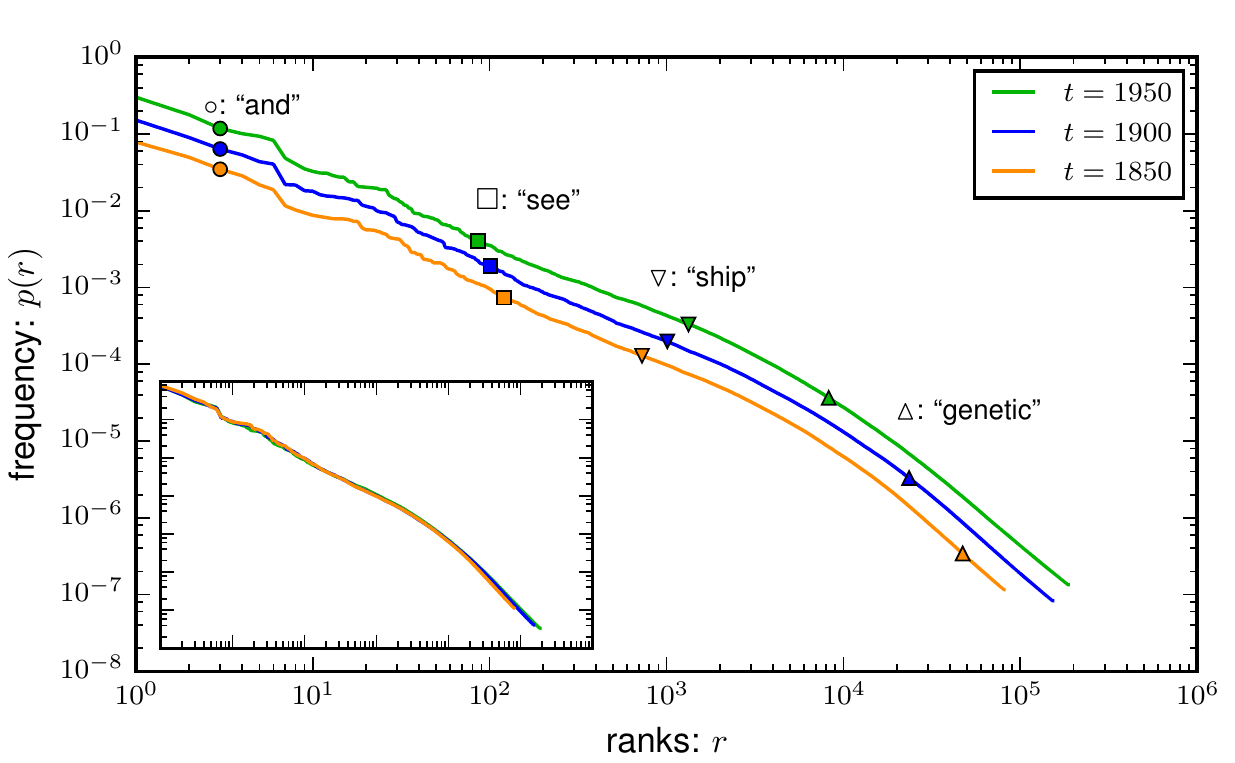}
\caption{
The English vocabulary in different years. Rank-frequency distribution $p(r)$ of
individual years $t$ for $t=1850$, $1900$, and $1950$ of the Google-ngram database~\cite{michel.2011}, multiplied by a
factor of $1$, $2$, and $4$, respectively, for better visual comparison. The inset shows the
original un-transformed data (same axis), highlighting that the rank-frequency
distributions are almost indistinguishable. Individual words
(e.g. ``and",``see",``ship",``genetic") show changes in rank and frequency (symbols),
where words with larger ranks (i.e. smaller frequencies) 
show larger change.     
}
\label{fig.changetime-jsd-intro}
\end{figure}

Quantifying the similarity  of symbolic sequences is a classical problem in information theory~\cite{kullback.book1959} with modern applications in linguistics~\cite{manning.book1999}, genetics~\cite{grosse.2002}, and image processing~\cite{he.2003}. 
The availability of large databases of texts sparked a renewed interest in the problem of similarity of the vocabulary of different collections of texts~\cite{boyack.2011,masucci.2011a,bochkarev.2014,pechenick.2015,cocho.2015}.
For instance, Fig.~\ref{fig.changetime-jsd-intro} shows the word-frequency distribution in
three large collections of English texts: from
$1850, 1900,$ and $1950$.  We see that the distribution itself remains essentially the same, a
heavy-tailed Zipf distribution~\cite{gerlach.2013}
\begin{equation}\label{eq.zipf}
p(r) \propto r^{-\gamma},
\end{equation}
where $p$ is the frequency of the $r$-th most frequent word and $\gamma\gtrapprox1$.  Changes are seen in the frequency~$p$ (or rank) of specific  words, e.g., {\it ship} lost and
{\it genetic} won popularity. Measures that quantify such changes are essential to answer questions 
such as: Is the
vocabulary from 1900 more similar to the one from 1850 or to the one  from 1950? How similar are two vocabularies (e.g., from different years)? Are the two finite-size
observations compatible with a finite sample of the same underlying
vocabulary?   How similar are the vocabulary of different authors or disciplines?    How fast
is the lexical change taking place?

Heavy-tailed and broad distributions of symbol-frequencies such as Eq.~(\ref{eq.zipf}) are typical in natural
languages~\cite{zipf.book1936,gerlach.2013,baayen.book2001,ferrer.2005a,serrano.2009,baek.2011,gerlach.2013} and appear also in the DNA (n-grams of base pairs for large
$n$)~\cite{mantegna.1994}, in gene expression~\cite{furusawa.2003}, and in music~\cite{serra.2012}.  The slow  decay observed in a broad range of frequencies implies that there is no typical frequency for words and
therefore relevant changes can occur in different ranges of the $p$-spectrum, from the
few large-frequency words all the way to the many low-frequency words.  
This imposes a challenge to define similarity measures that are able to account for this variability and that also yield accurate estimations based on finite-size observations.

In this paper we quantify the vocabulary similarity using a spectrum of measures
$D_\alpha$ based on the generalized entropy of order $\alpha$ ($D_{\alpha=1}$ recovers the usual Jensen-Shannon
divergence). We show how varying~$\alpha$ magnifies differences in the vocabulary at
different scales of the (heavy-tailed) frequency spectrum, thus providing
different information on the vocabulary change. We then compute the accuracy (bias) and precision (variance) of
estimations of $D_\alpha$ based on sequences of size $N$ and find that in heavy-tailed
distributions the convergence is much slower than in non-heavy-tailed distributions (it often scales as $1/N^\beta$ with $\beta<1$). 
Finally, we come back to the problem of comparing the English vocabulary in the last two
centuries in order to  illustrate the relevance of our general results.

\section{Definition}
\label{sec.definition}
Consider the probability distribution $\pp = (p_1,p_2, \ldots, p_S)$ of a random variable
over a discrete, countable set of symbols $i=1, \ldots, S$ (where later we include the
possibility for $S \rightarrow \infty$). 
From an information theory standpoint, a natural measure to quantify the difference between two such probability distributions $\pp$ and $\qq$ is the Jensen-Shannon divergence (JSD)~\cite{lin.1991}
\begin{equation}
\label{eq.jsd}
 D(\pp,\qq) = H\left(\frac{\pp +\qq}{2}\right) - \frac{1}{2} H(\pp) - \frac{1}{2}H(\qq),
\end{equation}
where $H$ is the Shannon entropy~\cite{cover.book2006}
\begin{equation}
\label{eq.h.shannon}
 H(\pp) = -\sum_i p_i \log p_i.
\end{equation}
This definition has several properties which are useful in the interpretation as a distance:
i) $D(\pp,\qq)\geq 0$ where the equality holds if and only if $\pp = \qq$;
ii)  $D(\pp,\qq) = D(\qq,\pp)$ (it is a symmetrized
Kullback-Leiber-divergence~\cite{lin.1991});
iii)  $\sqrt{D(\pp,\qq)}$ fulfills the triangle inequality and thus is a
metric~\cite{endres.2003};
and iv) $D(\pp,\qq)$ equals the mutual information of variables sampled from $\pp$ and
$\qq$~\cite{grosse.2002}, i.e., $D(\pp,\qq)$ equals the average amount of information in one randomly sampled word-token about  
which of the two distribution it was sampled from~\cite{montemurro.2010}.
The JSD is widely used in the statistical analysis of language~\cite{manning.book1999}, e.g. to automatically find individual documents that are (semantically) related~\cite{boyack.2011, masucci.2011a} or to track the rate of evolution in the lexical inventory of a language over historical time scales~\cite{bochkarev.2014,pechenick.2015}.

Here we also consider the generalization of JSD in which $H$ in Eq.~(\ref{eq.h.shannon}) is replaced by the generalized entropy of order $\alpha$~\cite{havrda.1967}
\begin{equation}
\label{eq.h.tsallis}
  H_{\alpha}(\pp) = \frac{1}{1-\alpha} \left(\sum_i p_i^{\alpha} - 1 \right),
\end{equation}
yielding a spectrum of divergence measures $D_\alpha$ parameterized by $\alpha$, first introduced in Ref.~\cite{burbea.1982}. The usual JSD is retrieved
for $\alpha=1$.
In (non-extensive) statistical mechanics, Eq.~\eqref{eq.h.tsallis} has been first proposed in Ref.~\cite{tsallis.1988} and $D_{\alpha}$ is sometimes called Jensen-Tsallis divergence.
While similar generalizations can be achieved with other formulations of generalized entropies such as the Renyi-entropy~\cite{renyi.1961,he.2003}, the corresponding divergences can become negative.
In contrast, $D_\alpha$ is strictly non-negative and it was recently shown that
$\sqrt{D_{\alpha}(\pp,\qq)}$ is a metric for any $\alpha \in (0,2]$~\cite{briet.2009}. For
heavy-tailed distributions, Eq.~(\ref{eq.zipf}), $H_{\alpha}<\infty$ for
$\alpha>1/\gamma$. 

We define a normalized version of $D_{\alpha}$ as
\begin{equation}\label{eq.norm}
\tilde{D}_{\alpha}(\pp,\qq) =
\frac{D_{\alpha}(\pp,\qq)}{D^{\max}_{\alpha}(\pp,\qq)} 
\end{equation} 
where 
\begin{equation}\label{eq.Dmax}
\begin{alignedat}{2}
D^{\max}_{\alpha}(\pp,\qq) =& \dfrac{2^{1-\alpha}-1}{2} \left( H_{\alpha} \left( \pp \right) + H_{\alpha} \left( \qq \right)+ \dfrac{2}{1-\alpha}\right)
\end{alignedat}
\end{equation}
is the maximum possible $D_\alpha$ between $\pp$ and $\qq$ obtained assuming that the the set of symbols in each distribution (i.e.,
the support of $\pp$ and $\qq$) are disjoint.
The main motivation for using the measure~(\ref{eq.norm}) is that
$\tilde{D}_{\alpha}(\pp,\qq) \in [0,1]$, while the range of admissible values of
$D_{\alpha}$ depends on $\alpha$. This allows for a meaningful comparison of the
divergences $\tilde{D}_\alpha(\pp,\qq)$ and $\tilde{D}_{\alpha'}(\pp,\qq)$ for $\alpha \neq
\alpha'$ and therefore also for the full spectrum of $\alpha$'s. In general, the metric
properties of $D_\alpha$ are not preserved by $\tilde{D}_\alpha$. An exception is the case
in which the rank-frequency distribution $p(r)$ underlying all $\pp$'s and $\qq$'s is invariant
(see Fig.~\ref{fig.changetime-jsd-intro}). Noting that Eq.~(\ref{eq.Dmax}) is independent of the symbols
we obtain that $D^{\max}_\alpha(\pp,\qq)$ is a constant for all $\pp$'s and $\qq$'s and therefore the metric property is preserved for $\tilde{D}_{\alpha}$.

\section{Interpretation}
\label{sec.interpretation}
\begin{figure}[b]
\includegraphics[width=1\columnwidth]{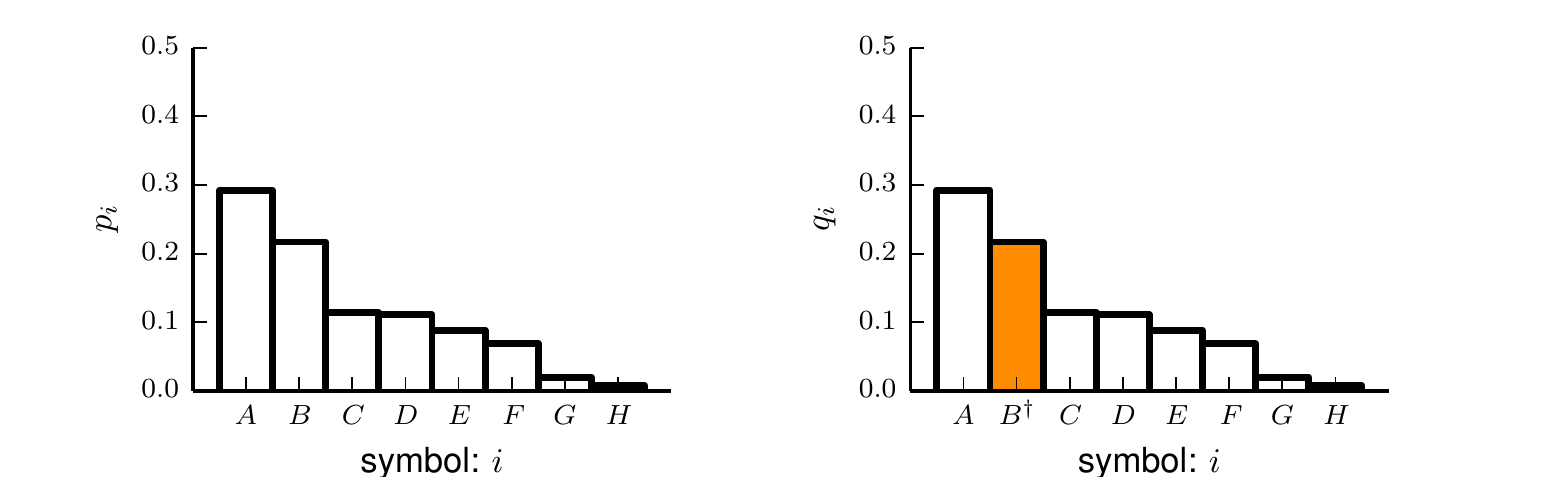}
\caption{
Illustration of our toy model where $\pp$ (left) and $\qq$ (right) have the same rank-frequency distribution, but differ in the probability for individual symbols. In this
example, $\pp$ and $\qq$ are the same ($p_i=q_i$) for $i\in \{A,C,D,E,F,G,H\}$, while the
symbol $i=B$ in $\pp$ is replaced by $i=B^{\dagger}$ in $\qq$ with  $p_{i=B}=q_{i=B^{\dagger}}$ and $p_{i=B^{\dagger}}=q_{i=B}=0$.
}
\label{fig.changetime-jsd-sketch}
\end{figure}

In order to clarify the interpretation of $D_\alpha$, it is useful to consider a toy model.
As in Fig.~\ref{fig.changetime-jsd-intro}, we consider  two
distributions $\pp$ and $\qq$ that have exactly the same rank-frequency 
distribution $p(r)$ but differ in (a subset of) the symbols they use. For simplicity, we
consider that symbols that differ in the two cases appear only in one of the
distributions. 
More precisely, denoting by $I_{p}=\{A,B,C,D,E,\ldots\}$ the set of symbols
in $\pp$ we define the set of replaced symbols as $I^* \subset I_p$.
The set of symbols in $\qq$ is chosen as $I_q=\{ i | i \in I_p \setminus I^*\} \cup \{ i^{\dagger} | i \in  I^*\}$ with probabilities $p_i=q_i$ for $i \in I_p \setminus I^*$ and $p_i=q_{i^{\dagger}}$ for $i \in I^*$, see Fig.~\ref{fig.changetime-jsd-sketch} for one example.

For a given distribution $\pp$ and a set of replaced symbols $I^*$, we compute
$D_{\alpha}(\pp,I^*)\equiv D_\alpha(\pp,\qq)$ as
\begin{equation}
\label{eq.jsd.t.model}
D_{\alpha}(\pp, I^*) = c_{\alpha}  \sum_{i\in I^*} p_i^\alpha,
\end{equation}
where $c_{\alpha}=(2^{(1-\alpha)}-1)/(1-\alpha)$. The maximum is given by
\begin{equation}
\label{eq.jsd.t.model.max}
D^{\max}_{\alpha}(\pp, I^*) = c_{\alpha}  \sum_{i\in I_p} p_i^\alpha
\end{equation}
such that 
\begin{equation}
\label{eq.jsd.t.model.rel}
 \tilde{D}_{\alpha}(\pp, I^*) = \frac{\sum_{i\in I^*} p_i^\alpha}{\sum_{i\in I_p} p_i^\alpha}.
\end{equation}
This shows that each symbol $i \in I^*$  that is replaced by a new symbol contributes $p_i^{\alpha}$ to $D_\alpha$.
It is thus clear that varying $\alpha$, the contribution of different frequencies become magnified (e.g. for $\alpha \gg 1$ large frequencies are enhanced while for $\alpha<0$ low frequencies contribute more to $D_{\alpha}$ than large frequencies).

In particular, for $\alpha=0$,
$\tilde{D}_{\alpha=0}(\pp, I^*)=\frac{|I^*|}{|I_p|}$ is the 
fraction of symbols (types) that are different in $\pp$ and $\qq$. Each symbol $i$ counts the same irrespective of their probabilities $p_i$.
For $\frac{|I^*|}{|I_p|}\ll 1$, $\tilde{D}_{\alpha=0}(\pp, I^*) = 1-J(I_p,I_q)$, where
$J(I_p,I_q) = \frac{|I_p \cap I_q|}{|I_p \cup I_q|}$ is the Jaccard-coefficient between
the two sets $I_p$ and $I_q$, an ad-hoc defined similarity measure widely used in information
retrieval~\cite{manning.book1999}.  For $\alpha=1$, $\tilde{D}_{\alpha=1}(\pp,
I^*)=\sum_{i \in I^*}p_i$ showing that  each replaced symbol is weighted by its
probability $p_i$ and thus that $\tilde{D}_{\alpha=1}$ measures the distance in terms of
tokens.  

\begin{figure}[bt]
\centering
\includegraphics[width=0.9\columnwidth]{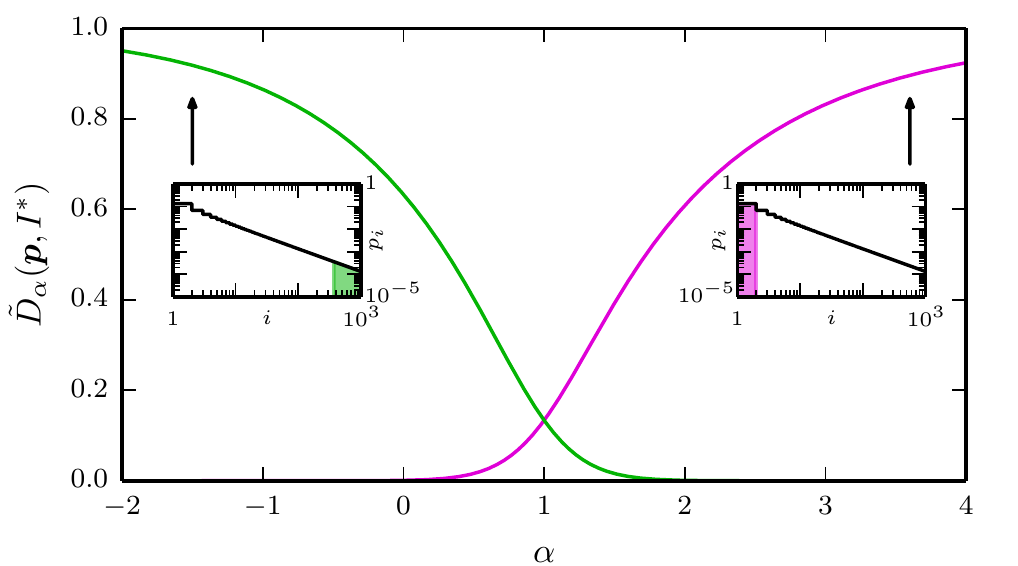}
\caption{The spectrum $\tilde{D}_\alpha(\pp,I^*)$ for two different changes.  The lines correspond to
  Eq.~(\ref{eq.jsd.t.model.rel}) with $p_i \propto i^{-1}$ with $i=1,2,\ldots,1000$ and
  two different sets of replaced symbols $I^*_1,I^*_2$.  Right inset: $I^*_1=\{1\}$, i.e.,
  only  the symbol with the highest probability, $p_{i=1} \approx 0.13$ is changed. 
Left inset: $I^*_2=\{368,\ldots,1000\}$, i.e, the symbols with small probability are
replaced. The choice of $I^*_2$ was made such that $\sum_{i \in I^*_2} p_i \approx
p_{i=1}$ and therefore   $\tilde{D}_{\alpha=1}(\pp, I^*_1) \approx \tilde{D}_{\alpha=1}(\pp, I^*_2)$.
}
\label{fig.changetime-jsd-model-Dq}
\end{figure}

The full spectrum $\tilde{D}_\alpha$ offers information on changes in all frequencies, a point which is particularly important for the case of heavy-tailed distributions because word-frequencies vary over many orders of magnitude.  
 Figure~\ref{fig.changetime-jsd-model-Dq} illustrates how different values of $\alpha$ are able to capture changes at different regions in the frequency-spectrum.
 In particular, it shows that $\tilde{D}_\alpha$ grows (decays) with $\alpha$ when the modified symbols have high (low) frequency. 
 Furthermore, the comparison between two given changes allow us to conclude about which change was more significant at different regions of the word-frequency spectrum. 
 In the example of the figure, both changes (the two lines) are equally significant from the point of view of the modified tokens ($\tilde{D}_1$ are the same), the change in the left affects more types ($\tilde{D}_0$ is larger), and the change in the right affects more frequent words ($\tilde{D}_\alpha$ is larger for $\alpha\gg 1$).

\section{Finite-size estimation}
\label{sec.finite}
In this section we turn to the estimation of $\tilde{D}_\alpha$ from data.  Even if $\tilde{D}_\alpha$
is defined with respect to distributions $\pp$ and $\qq$, Eq.~(\ref{eq.norm}), in practice these distributions are estimated from sequences
with finite-size $N$ (total number of symbols or word tokens) yielding finite size estimates of the distributions $\hat{\pp}$ and $\hat{\qq}$. 
The main obstacle in obtaining accurate estimates of $\tilde{D}_\alpha$  is that it requires the estimation of entropies for which, in general, unbiased estimators do not exist~\cite{schurmann.2004}.
Accordingly, even if $\pp=\qq$, in practice $H_\alpha(\hat{\pp}) \neq H_\alpha(\hat{\qq})$ and $\tilde{D}_\alpha(\hat{\pp},\hat{\qq})>0$ are measured
not only in single realizations, but also on average (the bias). Besides the bias, we are also interested in 
the expected fluctuation (standard deviation) of  the estimations of $H_\alpha$ and $\tilde{D}_\alpha$ and how both
they depend on the sequence size $N$ for large $N$. 
In heavy-tailed distributions such as Eq.~(\ref{eq.zipf}), these estimations are based on
an observed vocabulary $V$ (number of different symbols) that grows sub-linearly with $N$ as~\cite{herdan.book1960,heaps.book1978,gerlach.2014}
\begin{equation}\label{eq.heaps}
V(N) \propto N^{1/\gamma}.
\end{equation}
This implies that the entropies in Eq.~\eqref{eq.h.tsallis} are estimated based on a
sum of $V\rightarrow \infty$ terms (for $N\rightarrow \infty$). In practice, $\gamma$ and
the precise  functional form of the heavy-tailed distribution are unknown and therefore,
besides $\tilde{D}_\alpha$,  the estimation of $H_\alpha$ is also of interest (see
Ref.~\cite{dewit.1999} for the case in which a power-law form of $\pp$ is assumed to be
known a priori).

\subsection{Analytical Calculations}
Here we extend  previous
results~\cite{miller.book1955,basharin.1959,harris.1975,herzel.1994} and generalize them to arbitrary
$\alpha$. Given a probability distribution $\pp$ and the measured probabilities
$\hat{\pp}$ from a finite sample of $N$ word-tokens, we expand $H_{\alpha}(\hat{\pp})$ around the true probabilities $p_i$ up to second order as
\begin{equation}
\begin{alignedat}{2}
  H_{\alpha}(\hat{\pp}) \approx H_{\alpha}(\pp) +& \sum_{i: \hat{p}_i>0} (\hat{p}_i - p_i)\frac{\alpha}{1-\alpha}p_i^{\alpha-1}   \\
   -&\frac{1}{2} \sum_{i: \hat{p}_i>0} (\hat{p}_i-p_i)^2 \alpha p_i^{\alpha-2}
  \end{alignedat}
\end{equation}
where we used that $\frac{\partial H_{\alpha}}{\partial p_i} = \alpha/(1-\alpha)p_i^{\alpha-1}$ and $\frac{\partial^2 H}{\partial p_i \partial p_j} = -\alpha p_i^{\alpha-2}\delta_{i,j}$.
We then calculate $\mathbb{E}\left[ H_{\alpha}(\hat{\pp}) \right]$ by averaging over the
different realization of the random variables $\hat{p}_i$ by assuming that the absolute frequency of each symbol $i$
is drawn from an independent binomial with probability $p_i$ such that $\mathbb{E}\left[
  \hat{p}_i \right] = p_i$ and $\mathbb{V}\left[ \hat{p}_i \right] = p_i(1-p_i)/N \approx p_i/N$ yielding
\begin{equation}
\label{eq.jsd.halpha.bias.1}
\begin{alignedat}{2}
   \mathbb{E} \left[ H_{\alpha}(\hat{\pp}) \right]& \approx  H_{\alpha}(\pp) -
   \frac{\alpha}{2N} \sum_{i \in V} p_i^{\alpha-1} = H_{\alpha}(\pp) -  \frac{\alpha V^{(\alpha)}}{2N},
\end{alignedat}
\end{equation}
which defines the vocabulary size of order~$\alpha$
\begin{equation}
\label{eq.vocab.alpha}
 V^{(\alpha)} \equiv \sum_{i \in V} p_i^{\alpha-1}.
\end{equation}
From Eq.~(\ref{eq.jsd.halpha.bias.1}) we see that the bias in the entropy estimation
$|H_\alpha(\pp)-   \mathbb{E} \left[ H_{\alpha}(\hat{\pp}) \right]|$ depends only on
$V^{(\alpha)}$ and $N$. Similar calculations (see Appendix~\ref{appendixB})  show that the
large $N$ behavior of the bias and the fluctuations (variance) of $H_\alpha, D_\alpha,$ and $\tilde{D}_{\alpha}$
can be written as simple functions of $V^{(\alpha)}$ and $N$, as summarized in
Tab.~\ref{tab.1}. 
\begin{table}[h]
\begin{tabular}{c||c|c|c}
 & $H_{\alpha}$ & $D_{\alpha},\tilde{D}_{\alpha}(\bm{p}\neq\bm{q})$ & $D_{\alpha},\tilde{D}_{\alpha}(\bm{p}=\bm{q})$\tabularnewline
\hline 
Bias: & $V^{(\alpha)}/N$ & $V^{(\alpha)}/N$ & $V^{(\alpha)}/N$\tabularnewline
Fluctuations: & $V^{(2\alpha)}/N$ & $V^{(2\alpha)}/N$ & $V^{(2\alpha-1)}/N^{2}$\tabularnewline
\hline 
\end{tabular}
\caption{
Scaling of the bias $|\mathbb{E}[\hat{X}]-X|$ and  the fluctuations $\mathbb{V}[X]\equiv \mathbb{E}[\hat{X}^2]-\mathbb{E}[\hat{X}]^2$ of estimations $\hat{X}$.
  The results are valid for large sequence sizes $N$ and depend on the vocabulary of order $\alpha$, $V^{(\alpha)}$ as in Eqs.~(\ref{eq.vocab.alpha}) and~(\ref{eq.jsd.halpha.bias.2}).  
  Results are shown for $X=H_\alpha$  [order $\alpha$ entropy, Eq.~(\ref{eq.h.tsallis})], $D_\alpha$ [generalized divergence], $\tilde{D}_\alpha$  [normalized divergence, Eq.~(\ref{eq.norm})], see Appendix~\ref{appendixB} for the derivations. 
  For $\tilde{D}_\alpha$, we approximate $\tilde{D}_{\alpha} \approx D_{\alpha}/\mathbb{E}[D^{\max}_\alpha]$.
  }
  \label{tab.1}
\end{table}

\begin{table*}
\centering
\begin{tabular}{c||c|c|c|c|c}
 & $\mathbb{E}[H_{\alpha}(\hat{\bm{p}})]$ & \multicolumn{2}{c|}{$\mathbb{E}[D_{\alpha}(\hat{\bm{p}},\hat{\bm{q}})]$} & \multicolumn{2}{c}{$\mathbb{E}[\tilde{D}_{\alpha}(\hat{\bm{p}},\hat{\bm{q}})]$}\tabularnewline
\hline 
\hline 
$\alpha_{1}^{\mathbb{E}}$ & $1/\gamma$ & \multicolumn{2}{c|}{$1/\gamma$} & \multicolumn{2}{c}{$1/\gamma$}\tabularnewline
$\alpha_{2}^{\mathbb{E}}$ & $1+1/\gamma$ & \multicolumn{2}{c|}{$1+1/\gamma$} & \multicolumn{2}{c}{$1+1/\gamma$}\tabularnewline
\hline 
$\alpha<\alpha_{1}^{\mathbb{E}}$ & $cN^{-\alpha+1/\gamma}$ & \multicolumn{2}{c|}{$cN^{-\alpha+1/\gamma}$} & \multicolumn{2}{c}{$c$}\tabularnewline
$\alpha_{1}^{\mathbb{E}}<\alpha<\alpha_{2}^{\mathbb{E}}$ & $H_{\alpha}(\bm{p})+cN^{-\alpha+1/\gamma}$ & \multicolumn{2}{c|}{$D_{\alpha}(\bm{p},\bm{q})+cN^{-\alpha+1/\gamma}$} & \multicolumn{2}{c}{$\tilde{D}_{\alpha}(\bm{p},\bm{q})+cN^{-\alpha+1/\gamma}$}\tabularnewline
$\alpha>\alpha_{2}^{\mathbb{E}}$ & $H_{\alpha}(\bm{p})+cN^{-1}$ & \multicolumn{2}{c|}{$D_{\alpha}(\bm{p},\bm{q})+cN^{-1}$} & \multicolumn{2}{c}{$\tilde{D}_{\alpha}(\bm{p},\bm{q})+cN^{-1}$}\tabularnewline
\hline 
\multicolumn{1}{c}{} & \multicolumn{1}{c}{} & \multicolumn{2}{c}{} & \multicolumn{2}{c}{}\tabularnewline
 & $\mathbb{V}[H_{\alpha}(\hat{\bm{p}})]$ & \multicolumn{2}{c|}{$\mathbb{V}[D_{\alpha}(\hat{\bm{p}},\hat{\bm{q}})]$} & \multicolumn{2}{c}{$\mathbb{V}[\tilde{D}_{\alpha}(\hat{\bm{p}},\hat{\bm{q}})]$}\tabularnewline
 &  & $\bm{p}\neq\bm{q}$ & $\bm{p}=\bm{q}$ & $\bm{p}\neq\bm{q}$ & $\bm{p}=\bm{q}$\tabularnewline
\hline 
\hline 
$\alpha_{1}^{\mathbb{V}}$ & $1/(2\gamma)$ & $1/(2\gamma)$ & $1/(2\gamma)$ & $1/\gamma$ & $1/\gamma$\tabularnewline
$\alpha_{2}^{\mathbb{V}}$ & $\frac{1}{2}(1+1/\gamma)$ & $\frac{1}{2}(1+1/\gamma)$ & $1+1/(2\gamma)$ & $\frac{1}{2}(1+1/\gamma)$ & $1+1/(2\gamma)$\tabularnewline
\hline 
$\alpha<\alpha_{1}^{\mathbb{V}}$ & $cN^{-2\alpha+1/\gamma}$ & $cN^{-2\alpha+1/\gamma}$ & $cN^{-2\alpha+1/\gamma}$ & $cN^{-1/\gamma}$ & $cN^{-1/\gamma}$\tabularnewline
$\alpha_{1}^{\mathbb{V}}<\alpha<\alpha_{2}^{\mathbb{V}}$ & $cN^{-2\alpha+1/\gamma}$ & $cN^{-2\alpha+1/\gamma}$ & $cN^{-2\alpha+1/\gamma}$ & $cN^{-2\alpha+1/\gamma}$ & $cN^{-2\alpha+1/\gamma}$\tabularnewline
$\alpha>\alpha_{2}^{\mathbb{V}}$ & $cN^{-1}$ & $cN^{-1}$ & $cN^{-2}$ & $cN^{-1}$ & $cN^{-2}$\tabularnewline
\hline 
\end{tabular}
\caption{
Summary of finite size scaling for distributions with heavy tails. 
Mean ($\mathbb{E}$) and variance ($\mathbb{V}$) of the plug-in estimator of $H_{\alpha}$, $D_{\alpha}$, and $\tilde{D}_{\alpha}$ for samples $\hat{\pp}$ and $\hat{\qq}$ each of size $N$ drawn randomly from $\pp$ and $\qq$ with power-law rank-frequency distributions with exponent $\gamma>1$, Eq.~\eqref{eq.zipf}.
The results are obtained combining Tab.~\ref{tab.1} with Eq.~\eqref{eq.jsd.halpha.bias.2} (for details see Appendix~\ref{appendixB}, \ref{appendixC}).
The constant $c$ depends on $\alpha$ and has a different value in each case but is independent of $N$.
The limit $\gamma \rightarrow \infty$ corresponds to the case in which both $\pp$ and $\qq$ have short tails.
}
\label{tab.summary}
\end{table*}

We now focus on the dependence of $V^{(\alpha)}$ on $N$. The
sum $\sum_{i \in V}$ in Eq.~(\ref{eq.vocab.alpha}) indicates that in~$N$ samples, on 
average,  $V=V(N) \equiv V^{(\alpha=1)}$ different symbols are observed. If for $N\rightarrow \infty$ the
  vocabulary $V$ converges to a finite value, $V^{(\alpha)}$ in Eq.~(\ref{eq.vocab.alpha})
  also converges and the bias scales as $1/N$.  A more interesting scenario happens when
  $V$ grows with $N$. For the heavy-tailed case of interest here, $V$ grows as
  $N^{1/\gamma}$,  Eq.~(\ref{eq.heaps}), and  we obtain (in  Appendix~\ref{appendixC})  that 
  $V^{(\alpha)}$ scales for large $N$ as
\begin{equation}
\label{eq.jsd.halpha.bias.2}
 V^{(\alpha)} \propto \begin{cases}
N^{-\alpha+1+1/\gamma}, & \alpha<1+1/\gamma,\\
\text{ constant }, &  \alpha>1+1/\gamma,\\
\end{cases} 
\end{equation}
where $\gamma>1$ is the Zipf exponent defined in Eq.~(\ref{eq.zipf}) and $\alpha$  is the
order of the entropy in Eq.~(\ref{eq.h.tsallis}). 

From the combination of Eq.~(\ref{eq.jsd.halpha.bias.2}) and Tab.~\ref{tab.1} we obtain
the scalings with sequence size $N$ of the estimators of~$H_\alpha, D_\alpha,$ and $\tilde{D}_\alpha$ in a
heavy-tailed distribution with exponent~$\gamma$.
These scalings are summarized in Tab.~\ref{tab.summary}.
Three scaling regimes can be identified for the bias and for the fluctuations.  
(i) For large $\alpha$, the decay is $1/N$ (except when $\pp=\qq$, where the fluctuations decay even faster as $1/N^2$) as in the case of a finite vocabulary and short-tailed distributions.  
(ii) For intermediate $\alpha$, a sub-linear decay with $N$ is observed. 
This regime appears exclusively in heavy-tailed distributions and has important consequences in real applications, as shown below.  
From the exponents of the sub-linear decay we see that the bias decays more slowly than the fluctuations.  
(iii) For small $\alpha$, $\alpha<1/\gamma$, $H_\alpha(\pp)$ is not defined thus the estimator for the mean of $H_{\alpha}$ and $D_\alpha$ diverge.
The growth of $H_{\alpha}$ (and therefore $D_{\alpha}^{\max}$) and $D_{\alpha}$ with $N$ have the same scaling and therefore cancel each other for $\tilde{D}_\alpha$, in which case a convergence to a well defined value is found (the fluctuation of $\tilde{D}_\alpha$ still decays in this regime).

\begin{figure*}[btp]
\centering
\includegraphics[width=2\columnwidth]{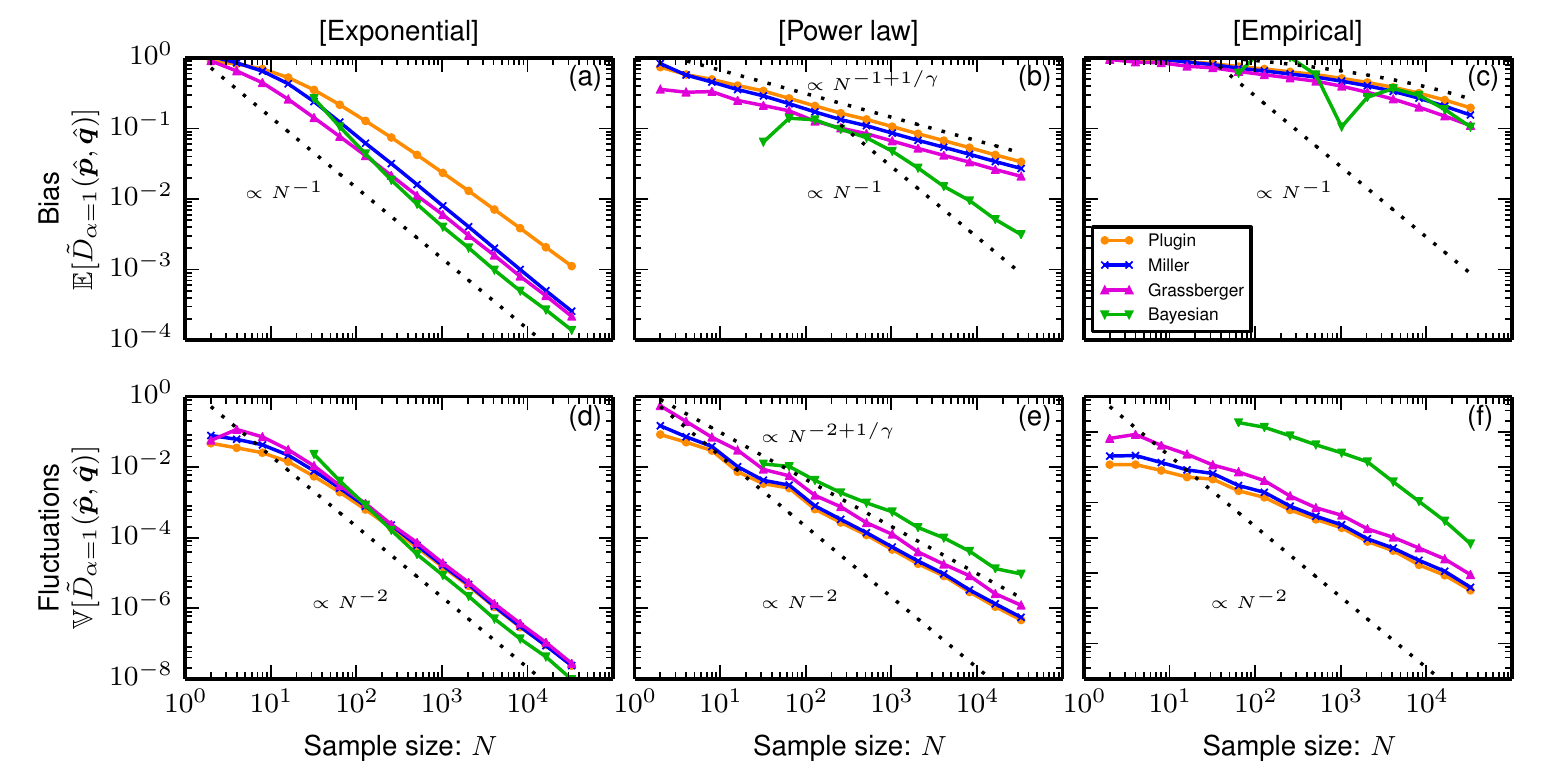}
\caption{
Finite-size estimation of the normalized Jensen-Shannon divergence~$\tilde{D}=\tilde{D}_{\alpha=1}$. 
(a-c) Estimation of $\mathbb{E}[\tilde{D}(\hat{\pp},\hat{\qq})]$ between two
sequences of size $N$  drawn from the same distribution (i.e. $ D(\pp,\qq)=0$) using four
different estimators of the entropy (see text) for three representative distributions: (a)
Exponential (short-tailed) distribution $p_i \propto e^{-a i}$ for $i=0,1,\ldots$ with
$a=0.1$; (b) Power-law (heavy-tailed) distribution $p_i \propto i^{-\gamma}$ for
$i=1,2,\ldots$ with $\gamma=3/2$; (c) Empirical Zipf-distribution of word frequencies, i.e. rank-frequency distribution $p(r)$ from the complete Google-ngram data, $p_i = f(i=r)$ for $i=1,\ldots,4623568$,  which is well described by a double power-law~\cite{gerlach.2013}. 
(d-f) Show the same as (a-c) for the fluctuations $\mathbb{V}[\tilde{D}(\hat{\pp},\hat{\qq})]$.
The dotted lines show the expected scalings from Tab.~\ref{tab.summary} for short-tailed distributions, i.e. $N^{-1}$ ($N^{-2}$), and power-law distributions, i.e. $N^{-1+1/\gamma}$ ($N^{-2+1/\gamma}$), for the bias (fluctuations). 
In (c) we show the expected scaling for the bias, $V_{\mathrm{emp}}(N)/N$, where $V_{\mathrm{emp}}(N)$ is the expected number of different symbols in a random sample of size $N$ from the empirical distribution~\cite{gerlach.2014}.
Averages are taken over $1000$ realizations. 
}
\label{fig.changetime-jsd-bias}
\end{figure*}

\subsection{Numerical Simulations}

Here we perform numerical estimations of the normalized divergence spectrum $\tilde{D}_\alpha$ that illustrate the regimes derived above, confirm
the validity of the approximations used in their derivations, and show that the same scalings are observed for
different entropy estimators. We sample twice $N$ symbols (tokens) from the same distribution
($\pp=\qq$),   and therefore $\tilde{D}_\alpha=0$ and the expected value $\mathbb{E}[\tilde{D}_{\alpha}(\hat{\pp},\hat{\qq})]$
is the bias. (The fact that the bias shows a slower decay with $N$ than the
  fluctuations  makes these two effects distinguishable also in this
$\tilde{D}_\alpha=0$ case because $\mathbb{E}[\tilde{D}_{\alpha}(\hat{\pp},\hat{\qq})] \gg \mathbb{V}[\tilde{D}_{\alpha}(\hat{\pp},\hat{\qq})]$ for large $N$).

\begin{figure*}[btp]
\centering
\includegraphics[width=1.99\columnwidth]{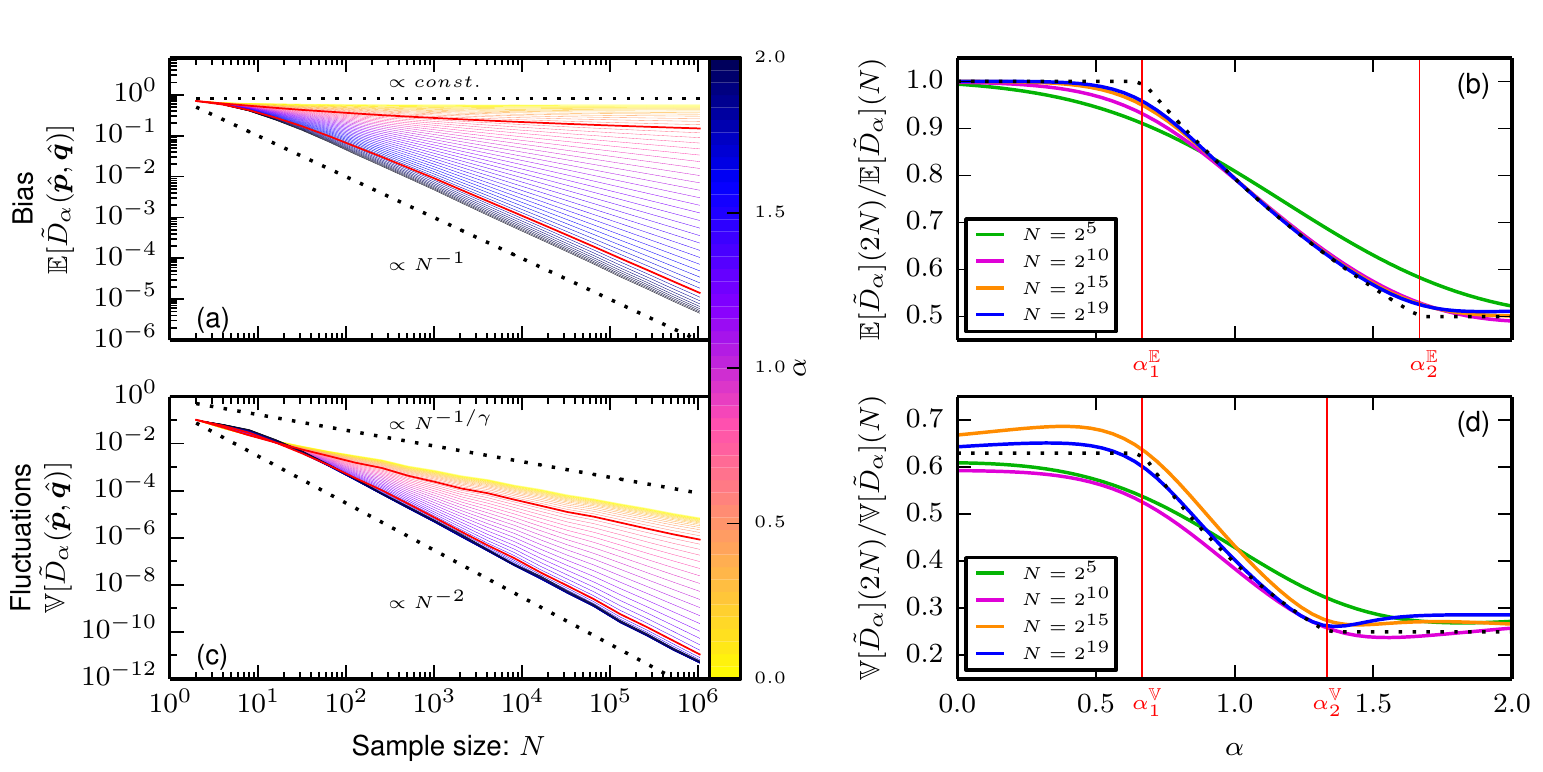}
\caption{
Bias (a,b) and fluctuations (c,d) in finite-size estimation of $\tilde{D}_\alpha$.
Estimation of $\mathbb{E}[\tilde{D}_{\alpha}(\hat{\pp},\hat{\qq})]$ between two sequences each of size $N$  drawn numerically from the same  power-law distribution $p_i \propto i^{-\gamma}$ for $i=1,2,\ldots,V\rightarrow \infty$ with $\gamma=3/2$  using the plugin-estimator ($p_i \mapsto \hat{p}_i$) for the entropies of order $\alpha$.
(a) Scaling of the bias with $N$ for different $\alpha$. 
(b) Decrease of the bias in $\tilde{D}_\alpha$ when sample size is doubled ($N \mapsto 2N$) for different values of $N$ as a function of $\alpha$. 
(c) and (d) show the same as (a) and (b) for the fluctuations $\mathbb{V}[\tilde{D}_{\alpha}(\hat{\pp},\hat{\qq})]$.
Red lines in all plots indicate the borders between the regimes, $\alpha^{\mathbb{E}}_1 = 1/\gamma =2/3$, $\alpha_2^{\mathbb{E}}=1+1/\gamma=5/3$ (for the bias in a,b), and $\alpha_1^{\mathbb{V}}=1/\gamma = 2/3$, $\alpha_2^{\mathbb{V}}=1+1/(2\gamma) = 4/3$ (for the fluctuations in c,d). 
Dotted lines indicate the predictions based on Tab.~\ref{tab.1} for $\alpha<\alpha_1^{\mathbb{E}},\alpha_1^{\mathbb{V}}$ and $\alpha>\alpha_2^{\mathbb{E}},\alpha_2^{\mathbb{V}}$ (in a,c) and all values of $\alpha$ (in b,d).
Averages are taken over $1000$ realizations.
}
\label{fig.changetime-jsd-alpha-bias}
\end{figure*}
We start with the most important prediction of our analytical calculations above: the existence in heavy-tailed distributions of a regime for which the bias and fluctuations of $\tilde{D}_\alpha$ decay with $N$ more slowly than $1/N$. 
This holds already for $\alpha=1$, i.e., for the usual Jensen-Shannon divergence, previously shown for the bias of $H_{\alpha=1}$ in Ref.~\cite{herzel.1994}. 
One potential limitation of our analytical calculations is that they are based on the plugin-estimator
obtained from replacing the $p_i$'s in the generalized entropies, Eq.~(\ref{eq.h.tsallis}), by the measured frequencies
(i.e. $p_i \mapsto \hat{p}_i=N_i/N$, with $N_i$ being the number of observed word tokens
of type $i$). To test the generality of our results, in the numerical simulations we use
four different estimators of the Shannon entropy (i.e., $\alpha=1$): i) the
\textit{Plugin}-estimator; ii) \textit{Miller}'s-estimator~\cite{miller.book1955}, which
takes into account the approximation obtained from the expansion in
Eq.~(\ref{eq.jsd.halpha.bias.1}); iii) \textit{Grassberger}'s
estimator~\cite{grassberger.2008}; and iv) a recently proposed \textit{Bayesian} estimator described
in~\cite{archer.2014} which is an extension of the approach by Nemenman et
al.~\cite{nemenman.2002} to the case where the number of possible symbols is unknown or
even countably infinite~\footnote{For small values of $N$, the Bayesian estimates of the individual entropies in some cases yield $\tilde{D}(\hat{\pp},\hat{\qq})<0$ such that we consider averages over $| \tilde{D}(\hat{\pp},\hat{\qq}) |$.}. The numerical results in Fig.~\ref{fig.changetime-jsd-bias} show
that the different estimators are indeed able to reduce the bias of the estimation, but
that the scaling of the bias with $N$ remains the same. In particular, the transition from short-tailed to heavy-tailed distribution leads to the predicted
transition from $N^{-1}$ ($N^{-2}$) to the slower $N^{-1+1/\gamma}$ ($N^{-2+1/\gamma}$) decay for the bias (fluctuations) for all estimators. The only exception is  in the bias of the Bayesian estimator for
the exact Zipf's law~(\ref{eq.zipf}),  but since this estimator shows a bad performance for the
fluctuation and for the real data we conclude that the slower scaling should be expected in
general also for this elaborated estimator. 
These results confirm the generality of our finding that the bias and fluctuation in $\tilde{D}_{\alpha=1}$ decays more slowly than $1/N$ in heavy-tailed distributions. 
The consequence of this result to applications will be discussed in the next section. 

We now consider the estimation of $\tilde{D}_\alpha$ for $\alpha \neq 1$ in the case of heavy-tailed distributions~(\ref{eq.zipf}). The numerical results in
Fig.~\ref{fig.changetime-jsd-alpha-bias} confirm the existence of the three scaling
regimes discussed after Eq.~(\ref{eq.jsd.halpha.bias.2}) and in Tab.~\ref{tab.summary}.  The panels (b) and (d) show the
relative reduction in the bias and fluctuations achieved when the sequence size is
doubled. For many different $\alpha$'s the relative reduction is larger than $0.5$ ($0.25$) for the bias (fluctuations), a
consequence of the slow decay with $N$ that shows the difficulty in obtaining a good
estimation of $\tilde{D}_\alpha$. In practice, the exponent
$\gamma$ of the distribution is unknown such that the critical values of $\alpha$ that
separates these regimes (e.g. $\alpha_1^{\mathbb{E}}=1/\gamma$ and $\alpha_2^{\mathbb{E}}=1+1/\gamma$ for the bias) cannot be
determined a priori. Yet,  since $\gamma>1$, we know that: (i) $\alpha_1^{\mathbb{E}},\alpha_1^{\mathbb{V}}< 1$  and
therefore $D_\alpha$ for $\alpha\ge1$ is such that $D_\alpha(\pp,\pp)=0$ for $N\rightarrow
\infty$; and  (ii) $\alpha_2^{\mathbb{E}},\alpha_2^{\mathbb{V}}<2$ and therefore the bias and fluctuations of $D_\alpha$
for $\alpha \ge 2$ decay as $1/N$ (or $1/N^2$ for the fluctuations in the case of $\pp=\qq$). This suggests $D_{\alpha=2}$ as a pragmatic choice for
empirical measurements because any further increase in $\alpha$ will not lead to a faster
convergence. 

\section{Application to real data}
\label{sec.application}
In this section, we show the significance of the general results of the previous section to specific problems. 

\begin{figure*}[bt]
\includegraphics[width=1\columnwidth]{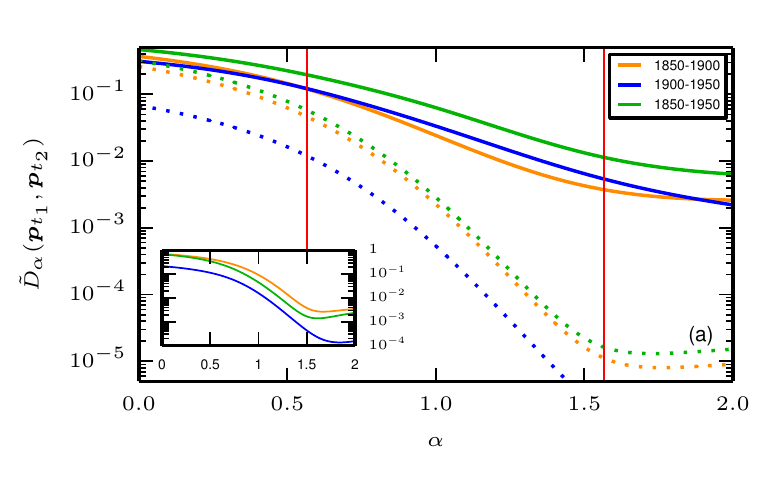}
\includegraphics[width=1\columnwidth]{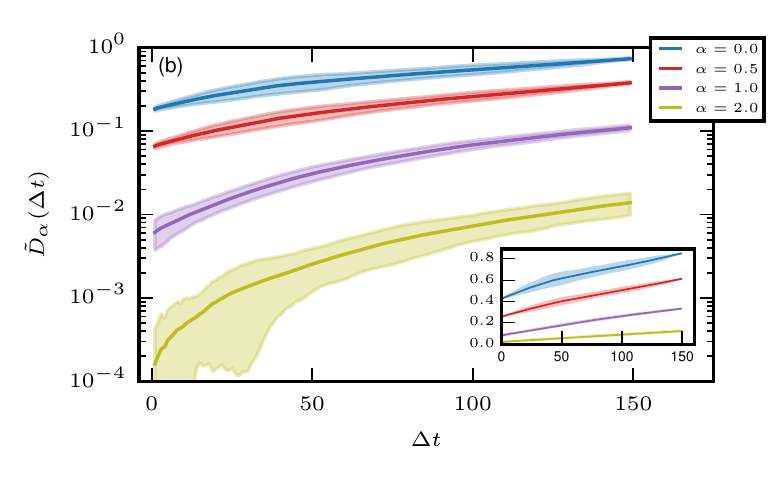}
\caption{
Measuring change in the usage of language on historical time-scales. 
(a) $\tilde{D}_{\alpha}(\pp_{t_1},\pp_{t_2})$ as a function of $\alpha$
for pairs of word-frequency distributions of the Google-ngram database obtained from
the yearly corpora $t_1$ and $t_2$ with $(t_1,t_2) \in \{(1850,1900),(1900,1950),(1850,1950)\}$ (solid lines). 
The dotted lines with the same colors show the results of a null model in which samples of the same size of the ones in $t_1$ and $t_2$ are randomly drawn from the {\it same} distribution (obtained from combining the corpora in $t_1$ and $t_2$) mimicking a minimum distance that can be observed due to finite-size effects.
The vertical lines show the three regimes $\alpha<1/\gamma$, $1/\gamma<\alpha<1+1/\gamma$, and $\alpha>1+1/\gamma$ in the convergence of $\tilde{D}_{\alpha}(\pp_{t_1},\pp_{t_2})$ with $N$ (see Sec.~\ref{sec.finite}), obtained using $\gamma=1.77$~\cite{gerlach.2013}.
Inset: ratio $\tilde{D}_{\alpha}(\pp_{t_{12}},\pp_{t_{12}})/\tilde{D}_{\alpha}(\pp_{t_{1}},\pp_{t_{2}})$.
(b) Average divergence as a function of $\Delta t \equiv | t_2- t_1|$, calculated as
$\tilde{D}_{\alpha}(\Delta t) = \frac{1}{N_{\Delta t}}\sum_{t_1=1805}^{2000-\Delta t}
\tilde{D}_{\alpha}(\pp_{t_1},\pp_{t_1+\Delta t})$ for four different $\alpha$ (solid lines).
Shaded areas represent the standard deviation associated to the average $\tilde{D}_{\alpha}(\Delta t)$.
Inset: $\sqrt{\tilde{D}_{\alpha}(\Delta t)}$ as a function of $\Delta t$ highlighting the approximate relationship $\tilde{D}_{\alpha}(\Delta t) \sim \Delta t^2$ for $\Delta t \gg 1$. 
}
\label{fig.changetime-jsd-app}
\end{figure*}
A problem that appears in different contexts is to test whether two finite-size $N$ sequences, described by their
empirical distributions $\hat{\pp}$ and $\hat{\qq}$,  have a common source (null
hypothesis). This involves the computation of a single divergence
$\tilde{D}_{\alpha}(\pp,\qq)$, which is then compared to the divergence
$\tilde{D}_{\alpha}(\pp',\pp')$ between two
finite-size (random) samplings of a single (properly chosen) distribution~$\pp'$ (e.g.,
$\pp'=0.5 \pp+0.5 \qq$). 
The probability of observing $ \tilde{D}_{\alpha}(\pp',\pp') \ge
\tilde{D}_{\alpha}(\pp,\qq) $ is then reported as
a p-value~\cite{grosse.2002}.  Besides applications in language, e.g. comparing the distribution of word-frequencies, this problem appears in the identification of coding- and non-coding regions in DNA~\cite{bernaola.2000}.  
The significance of our results for finite-size  estimations in Sec.~\ref{sec.finite} is
that for the case of  heavy-tailed distribution the expected $\tilde{D}_{\alpha}(\pp,\qq)$ of the
null model may be  much larger than the predicted value based on a $1/N$ decay (as observed in short-tailed distributions).  
If the slower convergence in $N$ is ignored, e.g., by applying standard
tests~\cite{grosse.2002} to heavy-tailed distributions,  one rejects the null hypothesis (low p-value) even if the data is drawn from the same source because the measured distance will be much larger. 
The example in Fig.~\ref{fig.changetime-jsd-bias}(c) shows that, even when the size of both sequences is on the order of $N \approx 10^5$, the expected $\tilde{D}_1$ (JSD) is
$\mathbb{E}[\tilde{D}_{\alpha=1}(\hat{\pp},\hat{\qq})]\approx 10^{-1}$. 
This is two orders of magnitude larger than for the exponential distribution in Fig.~\ref{fig.changetime-jsd-bias}(a), where  $\mathbb{E}[\tilde{D}_{\alpha=1}(\hat{\pp},\hat{\qq})] \approx 10^{-3}$.

The next problems we consider appear in the analysis of historical data and in the quantification of
language change~\footnote{We use the Google-ngram corpus containing the frequency of usage
  of millions of words from 4\% of all books ever published in the English language with a
  yearly resolution~\cite{michel.2011}.}.
 These problems are representative of problems
that  involve the comparison of two or more divergences
$\tilde{D}_{\alpha}(\pp,\qq)$, obtained from different distributions $\pp \ne \qq$
and $\alpha'\neq \alpha$. 
As depicted in Fig.~\ref{fig.changetime-jsd-intro}, the different distributions are
obtained based on individual years ($t \in \{1850,1900,1950\}$) and  we calculate the normalized spectrum
$\tilde{D}_{\alpha}(\pp_{t_1},\pp_{t_2})$ between pairs of years $(t_1,t_2)$. 
As argued in Sec.~\ref{sec.definition} and Appendix~\ref{appendixA},
$\tilde{D}_{\alpha}(\pp,\qq)$ is meaningful even if the sequences used to estimate $\pp$
and $\qq$ have different sizes~$N_p\neq N_q$. We summarize our results in
Fig.~\ref{fig.changetime-jsd-app}, from which different conclusions can be drawn:

\paragraph{Temporal change.}  The change of English from $1850$ to $1950$ was larger
than the change form $1850$ to $1900$  and from $1900$ to $1950$, as seen from the fact
that the curve of $\tilde{D}_{\alpha}(\pp_{1850},\pp_{1950})$ in Fig.~\ref{fig.changetime-jsd-app}a lies
above the two other curves for all $\alpha$. This intuitive result (evolutionary dynamics
show no recurrences) confirms that the divergence spectrum
$\tilde{D}_{\alpha}(\pp_{t_1},\pp_{t_2})$ is a meaningful quantification
of language change. The average dependency of $\tilde{D}_{\alpha}(\pp_{1850},\pp_{1950})$ on
$\Delta t = |t_2 - t_1 |$, shown in  Fig.~\ref{fig.changetime-jsd-app}b, can be
thus used as a quantification of  the speed of language change.
We observe an approximate relationship $\tilde{D}_{\alpha}(\Delta t) \approx \tilde{D}_{\alpha}^{(i)}+\tilde{D}_{\alpha}^{(ii)}\Delta t^2$ for $\Delta t \gg 1$ (see Inset Fig.~\ref{fig.changetime-jsd-app}b), where $\tilde{D}_{\alpha}^{(i)}$ and $\tilde{D}_{\alpha}^{(ii)}$ are constants and can be related to words that change due to fluctuations (finite sampling or topical dependencies) which is independent of $\Delta t$ and words that show a systematic increase or decrease over $\Delta t$, respectively (see Appendix \ref{appendixD} for a detailed discussion).

\paragraph{Dependence on $\alpha$.} All observed divergences
$\tilde{D}_{\alpha}(\pp_{t_1},\pp_{t_2})$  decay with $\alpha$ (e.g., the three curves in
Fig.~\ref{fig.changetime-jsd-app}a). As discussed in Sec.~\ref{sec.interpretation}, this
shows that for words with a high (low) frequency the distributions are more (less) similar
and thus the change is slower (faster).  
This result is consistent with previous works on the evolution of individual words on historical time scales reporting that frequent words tend to be more stable~\cite{pagel.2007,lieberman.2007}.
This dependence on $\alpha$  is essential when
comparing the change $1850 \mapsto 1900$ to the change $1900 \mapsto 1950$
(Fig.~\ref{fig.changetime-jsd-app}a). While the earlier change was smaller if counted on a
token basis, $\tilde{D}_{\alpha=1}(\pp_{1850},\pp_{1900}) <
\tilde{D}_{\alpha=1}(\pp_{1900},\pp_{1950})$, it becomes larger if one focus on the more
frequent words $[\tilde{D}_{\alpha=2}(\pp_{1850},\pp_{1900}) >
\tilde{D}_{\alpha=2}(\pp_{1900},\pp_{1950})$].

\paragraph{Role of finite-size scalings.} Our finding that the scalings (of the bias and
of the fluctuations) in
$\tilde{D}_\alpha$ with sample size $N$ depend on $\alpha$ allows for a deeper
understanding of the $\tilde{D}_{\alpha}(\pp_{t_1},\pp_{t_2})$  measurements discussed above. The expected
$\tilde{D}_\alpha$'s for random sampling of the same distribution
(null model shown as dashed line in Fig.~\ref{fig.changetime-jsd-app}a)  are of the same
order as the empirical distance for small $\alpha$ (i.e. $\tilde{D}_{\alpha}(\pp_{t_{12}},\pp_{t_{12}}) \approx
\tilde{D}_{\alpha}(\pp_{t_{1}},\pp_{t_{2}})$) and it is only for $\alpha>1$ that the null 
model divergence becomes negligible compared to the empirical divergence (i.e. $\tilde{D}_{\alpha}(\pp_{t_{12}},\pp_{t_{12}}) \ll
\tilde{D}_{\alpha}(\pp_{t_{1}},\pp_{t_{2}})$).  This implies that even though the size of
the individual corpora is of the order of $N \approx 10^9$ word-tokens, the empirically
measured $\tilde{D}_{\alpha}$ is still strongly influenced by finite-size effects
over a wide range of values for $\alpha$, in agreement with our analysis in
Sec.~\ref{sec.finite}. In particular, the bias for Jensen-Shanon
  divergence ($\alpha=1$) is  important even for the case of 
  the  (extremely large)  Google-ngram database (e.g., the Inset of
  Fig. \ref{fig.changetime-jsd-app}a shows that the bias is $\approx 10\%$).

\paragraph{$\alpha=2$ as a pragmatic choice.}
The slow decay of bias and fluctuations with database size suggests that $\tilde{D}_{\alpha=2}$ is a
pragmatic choice in reducing such 
finite-size effects when the exponent~$\gamma$ in the power-law distribution is not known. 
This conclusion is further corroborated in the analysis of the dependence of
$\tilde{D}_{\alpha}$ with $\Delta t$ (Fig.~\ref{fig.changetime-jsd-app}b). 
While $\tilde{D}_{\alpha}(\Delta t = 0)=0$ by construction, $\tilde{D}_{\alpha}$  does not converge to zero for  $\Delta t \rightarrow 0$ when extrapolating from $\tilde{D}_{\alpha}(\Delta t>0)$,
but instead it seems to saturate, i.e. $\tilde{D}_{\alpha}(\Delta t \rightarrow 0)\approx \tilde{D}^{(i)}_{\alpha}>0$.
For small values of $\alpha$, $\tilde{D}^{(i)}_{\alpha}$ is of the same order of magnitude of the expected bias (e.g.,  shown as dashed line in
Fig.~\ref{fig.changetime-jsd-app}a) and even of the same order of magnitude of the divergence  $\tilde{D}_{\alpha}(\Delta t=100)$ between two corpora
separated by $100$ years. For small $\alpha$ and $\Delta t$,  it is thus difficult to
distinguish between finite-size effects ($\tilde{D}^{(i)}_{\alpha}$) and actual language change. 
 Results for $\alpha=2$ show the largest relative variation with $\Delta t$ and are therefore statistically more suited to  
quantify language change over time. 

\section{Conclusions}
\label{sec.conclusions}
In summary, we investigated the use of generalized entropies $H_\alpha$ to quantify the difference between symbolic sequences with heavy-tailed frequency distributions.
In particular, we introduced a normalized spectrum of a generalized divergence, $\tilde{D}_\alpha(\pp,\qq)$  in Eq.~(\ref{eq.norm}), that allows for a comparison between the different distributions $\pp$ and $\qq$ and also for different $\alpha$'s.  
Increasing $\alpha$, $\tilde{D}_\alpha$ attributes higher weights to high-frequency symbols. 
The more complete characterization given by the full spectrum $\tilde{D}_\alpha$ is particularly important in the case of heavy-tailed distributions because in this case symbols do not have a characteristic frequency but instead show frequencies on a broad range of values. 

Our main analytical finding is how the systematic (bias) and statistical (fluctuations) errors of finite-size ($N$) estimations of $H_\alpha$ and $\tilde{D}_\alpha$ scales with $N$, see  Tab.~\ref{tab.summary}.  The existence of regimes in which these scalings decay slower than $1/N$ shows that large uncertainties should be expected in $H_\alpha$ and $\tilde{D}_\alpha$ estimated even for very large databases. 
This should be  taken into account when comparing two or more $\tilde{D}_\alpha$'s and when estimating the probability of  two sequences  having the same source. 
The fact that for large $\alpha$ we recover the usual scaling (decay with $1/N$) suggests
$\tilde{D}_{\alpha=2}$ as a pragmatic choice in applications involving heavy-tailed
distributions. Previous works using information theoretic measures in language used
  $\alpha=1$~\cite{boyack.2011,masucci.2011a,bochkarev.2014,pechenick.2015} and did not
  take into account the effect of (finite) database size.
Our results show that the bias and fluctuations are significant even in the extremely
large Google-ngram database. It is therefore essential to clarify what is the role of
finite-size effects in the reported conclusions, in particular in the (typical) case that
database sizes change over time.

Our main empirical findings on language change are: 
i) that least frequent words contribute more to the total vocabulary change; 
ii) the answer to the question whether the speed of language change is  accelerating depends on the emphasizes that is given to either low-frequency or high-frequency words; and iii) the quantification of the speed of vocabulary change in time, $\Delta t$,  which shows roughly a dependence $\tilde{D}_{\alpha}(\Delta t)  \approx \tilde{D}_{\alpha}^{(i)}+\tilde{D}_{\alpha}^{(ii)}\Delta t^2$, where $\tilde{D}_{\alpha}^{(i)}$ ($\tilde{D}_{\alpha}^{(ii)}$) quantifies the degree to which words change due to fluctuations independent of time (systematic increase/decrease of the frequency over time).
More generally, our spectrum $\tilde{D}_{\alpha}$ opens the possibility of studying language change at different resolution, combining aspects from the analysis on the level of individual words (e.g. Refs.~\cite{pagel.2007,lieberman.2007}) and the full vocabulary of a language (e.g. Refs.~\cite{bochkarev.2014,cocho.2015}).

Our results are also of interest beyond the cases treated here.
First, the finite-size scaling we derive appear already in the entropy and therefore the same scalings are expected in any entropy-based measure, including those based on conditional entropies such as the Kullback-Leibler divergence~\cite{kullback.book1959}.
Second, the analysis is not necessarily restricted to the word level, it can be straightforward extended also to n-grams of words which also show heavy-tailed distributions~\cite{yasseri.2012}. 
Third, the spectrum of divergences $\tilde{D}_\alpha(\pp,\qq)$ offers a unifying framework which can be applied to problems involving different partitions of texts by varying the parameter $\alpha$.
For example, while in document classification~\cite{manning.book1999} one tries to identify topical words (suggesting the use of low values of $\alpha$), in applications of authorship attribution~\cite{stamatatos.2009} it has been shown that the comparison of the most-frequent (function) words yields the best results (suggesting the use of large values of $\alpha$).
Fourth, heavy-tailed distributions appear
in different problems involving symbolic sequences (e.g., in the DNA
~\cite{mantegna.1994}, in gene expression~\cite{furusawa.2003}, and in
music~\cite{serra.2012}), and the significance of our results is that they can be applied
in all these cases.

\acknowledgments
We thank Peter Grassberger for insightful discussions.

\appendix
\section{Documents with different lengths}
\label{appendixA}

Here we discuss how to proceed if the JSD is computed from finite datasets with different
finite lengths $N$, i.e. when $\pp$ ($\qq$) is estimated from a
sequence of length $N_p$ ($N_q\neq N_p$).

\subsection{Different Weights}

A possible way to extend Eq.~\eqref{eq.jsd} taking into account the unequal contribution $N_p \neq N_q$ is to consider weights~$\pi$ as~\cite{grosse.2002}
\begin{equation}
\label{eq.jsd.tsallis2}
 D^{\pi}_{\alpha}(\pp,\qq) = H_{\alpha}(\pi_p\pp +\pi_q\qq) - \pi_p H_{\alpha}(\pp) - \pi_qH_{\alpha}(\qq).
\end{equation}
with $\pi_p=N_p/N$ and $N_q/N$ such that $\pi_p+\pi_q=1$ with $N=N_p+N_q$ (denoted as natural weights in the following). 
Obviously, if $N_p=N_q$ then $\pi_p=\pi_q=1/2$ and $D_\alpha$ is recovered. 
The normalized distance~(\ref{eq.norm}) becomes
\begin{equation}\label{eq.norm2}
\tilde{D}^\pi_{\alpha}(\pp,\qq) =
\frac{D^\pi_{\alpha}(\pp,\qq)}{D^{\pi,\max}_{\alpha}(\pp,\qq)} ,
\end{equation} 
where
\begin{equation}\label{eq.Dmax2}
\begin{alignedat}{2}
D^{\pi,\max}_{\alpha}(\pp,\qq)=& \left( \pi_p^{\alpha} - \pi_p \right) H_{\alpha} \left( \pp \right) + \left( \pi_q^{\alpha} - \pi_q \right) H_{\alpha} \left( \qq \right)\\
&+ \frac{1}{1-\alpha} \left( \pi_p^{\alpha} + \pi_q^{\alpha} - 1 \right).
\end{alignedat}
\end{equation}
Our main results for the finite-size scaling of $D_\alpha$ summarized in Tab.~\ref{tab.summary} remain valid for the weighted divergences.

The approach above follows Ref~\cite{grosse.2002}, which introduced weights to the usual JSD (non-normalized, $\alpha=1$) and showed that the natural weights $\pi_p=N_p/N$ and $\pi_q=N_q/N$ imply certain useful properties for the JSD, e.g., that the bias does not depend on the relative size of the two samples.
While their main motivation was to compare the statistical significance of a single measurement of the JSD in the identification of stationary subsequences (of possibly different lengths) in a non-stationary symbolic sequence, here, we are mainly interested in comparing two (or more) measured distances.
In this case, choosing weights that depend on the size of the individual samples becomes problematic when the sequences are of different lengths.
The demonstration that $\sqrt{D_{\alpha}(\pp,\qq)}$ is a metric for any $\alpha \in (0,2]$~\cite{briet.2009} is valid for fixed weights $\pi_p=\pi_q=1/2$. 
More generally, the measure $D_\alpha^\pi$ itself depends on the weights $\pi$ such that $D_\alpha^\pi$ and $D_\alpha^{\pi'}$ constitute different measures when $\pi \neq \pi'$.
It is therefore not meaningful to compare $D_\alpha^\pi(\pp,\qq)$ and $D_\alpha^{\pi'}(\pp',\qq')$ if $N_p/N_q \neq N_{p'}/N_{q'}$ because this would imply that $\pi' \neq \pi$.

\subsection{Equal Weights}
In the previous section we argued that it is essential to choose fixed weights $\pi$ when comparing different distances.
The choice of equal weights $\pi_p=\pi_q=1/2$ can, however, still be interpreted in the framework of natural weights ($\pi_p=N_p/N$, $\pi_q=N_q/N$) as the distance between undersampled versions of the sequences.
For given $\pp$ and $\qq$ with $N_p \neq N_q$ we choose equal weights $\pi_p=\pi_q=1/2$ yielding a distance $D^{1/2}_{\alpha}(\pp,\qq)$.
If we randomly draw samples $\pp'$ and $\qq'$ of size $N_p'=N_q'$ from the distributions $\pp$ and $\qq$, (by construction) the natural weights coincide with the equal weights, i.e. $\pi_p'=\pi_q'=N_p'/N=N_q'/N=1/2$, and $\underset{N_p'=N_q'\rightarrow \infty}{\lim} D^{\pi'}_{\alpha}(\pp',\qq')=D^{1/2}_{\alpha}(\pp,\qq)$.

\begin{figure}
\includegraphics[width=1\columnwidth]{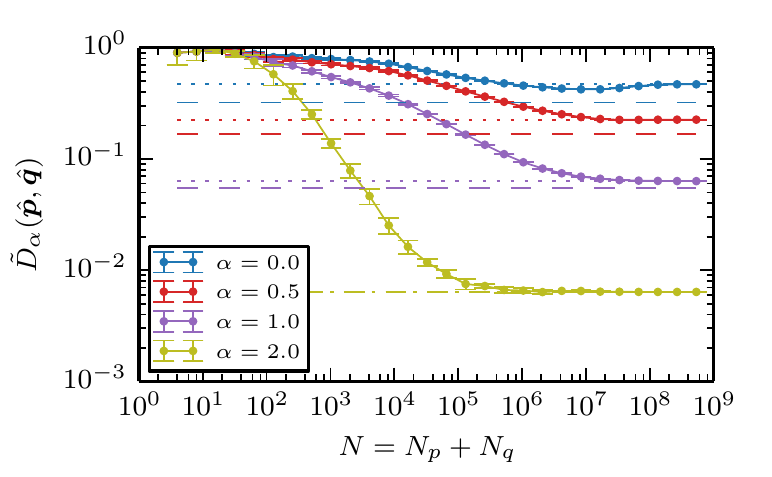}
\caption{
JSD-$\alpha$ for sequences of different lengths. Measurement of $\tilde{D}_{\alpha}(\hat{\pp},\hat{\qq})$ between sequences $\hat{\pp}$, $\hat{\qq}$ of size $N_p'=N_q'$ sampled randomly from the empirical distribution of the Google-ngram of the years $t \in \{1850,1950\}$ with different sizes, i.e. $\pp=\pp_{t=1850}$ and $\qq=\pp_{t=1950}$ with $N_p \neq N_q$, as a function of the sample size $N'=N_p'+N_q'$ for different values of $\alpha$.
The dotted (dashed) lines show $\tilde{D}^{\pi}_{\alpha}(\pp,\qq)$ between the full distributions $\pp$ and $\qq$ with equal (natural) weights, i.e. $\pi_p=\pi_q=1/2$ ($\pi_p=N_p/(N_p+N_q)\approx 0.22$ and $\pi_q=N_q/(N_p+N_q)\approx 0.78$ corresponding to the relative size of $\pp$ and $\qq$)
}
\label{fig.changetime-jsd-undersampling}
\end{figure}

In Fig.~\ref{fig.changetime-jsd-undersampling} we show the difference in $\tilde{D}^{\pi}_{\alpha}(\pp,\qq)$ between two empirical distributions from the Google-ngram with different sizes ($N_p \neq N_q$) when choosing equal and natural weights. 
Using equal weights corresponds to the case in which we draw samples $\hat{\pp}$ and $\hat{\qq}$ that are of equal length ($N_p'=N_q'$) such that equal and natural weights coincide and taking the limit $N_p',N_q'\rightarrow \infty$.

\section{Finite size estimation of $H_{\alpha}$, $D_{\alpha}$, and $\tilde{D}_{\alpha}$}
\label{appendixB}
In this section we present the calculations on the mean (i.e. the bias) and the fluctuations in finite-size estimates of $H_{\alpha}$, $D_{\alpha}$, and $\tilde{D}_{\alpha}$.
The starting point is a finite sample $\hat{\pp}=(n_1/N,n_2/N,\ldots,n_V/N)$ of size $N$ (where $n_i$ is the number of times symbol $i$ was observed) which we assume is obtained from $N$ identical and independent draws from the distribution $\pp$ giving an estimator for $H_{\alpha}$:
\begin{equation}
\label{eq.Halpha.emp}
 H_{\alpha}(\hat{\pp}) = \frac{1}{1-\alpha} \left(\sum_{i: \hat{p}_i>0} \hat{p}_i^{\alpha} -1 \right).
\end{equation}
In order to take the corresponding expectation values we expand $\hat{p}_i^{\alpha}$ around the true probabilities $p_i$ up to second order
\begin{equation}
\hat{p}_i^{\alpha} \approx p_i^{\alpha} + (\hat{p}_i - p_i) \alpha p_i^{\alpha-1} + \frac{1}{2} (\hat{p}_i - p_i)^2 \alpha (\alpha-1) p_i^{\alpha-2}\\
\end{equation}
and average over the realizations of the random variables $\hat{p}_i^{\alpha}$ by assuming that each symbol is drawn independently from binomial with probability $p_i$ such that $\langle (\hat{p}_i - p_i) \rangle = 0$ and $\langle (\hat{p}_i - p_i)^2 \rangle = p_i(1-p_i)/N \approx p_i/N$ yielding~\cite{herzel.1994}
\begin{equation}
\label{eq.pialpha.avg}
 \langle \hat{p}_i^{\alpha} \rangle \approx p_i^{\alpha} + \frac{1}{2N}\alpha (\alpha -1)p_i^{\alpha-1}.
\end{equation}

\subsection{$H_{\alpha}$}
Combining Eqs.~\eqref{eq.Halpha.emp} and \eqref{eq.pialpha.avg} we obtain for the mean
\begin{equation}
\begin{alignedat}{2}
 \mathbb{E}[H_{\alpha}(\hat{\pp})] &\equiv \langle H_{\alpha}(\hat{\pp}) \rangle = \frac{1}{1-\alpha} \left(\sum_{i \in \langle  V_{\hat{\pp}} \rangle} \langle \hat{p}_i^{\alpha} \rangle -1 \right)\\
 &= \frac{1}{1-\alpha} \left(\sum_{i \in \langle  V_{\hat{\pp}} \rangle} p_i^{\alpha}-1 \right) - \frac{\alpha}{2N}\sum_{i \in \langle  V_{\hat{\pp}} \rangle} p_i^{\alpha-1}\\
 &=  \frac{1}{1-\alpha} \left(V^{(\alpha+1)}_{\hat{\pp}}-1 \right) - \frac{\alpha}{2N} V^{(\alpha)}_{\hat{\pp}}
 \end{alignedat}
\end{equation}
where we introduce the notation $\sum_{i \in \langle  V_{\hat{\pp}} \rangle}$ indicating that we average only over the expected number of observed symbols $\langle V_{\hat{\pp}} \rangle$ in samples $\hat{\pp}$.

For the variance we get
\begin{equation}
\begin{alignedat}{2}
 \mathbb{V}[H_{\alpha}(\hat{\pp})] \equiv &  \mathbb{E}[H_{\alpha}(\hat{\pp})^2] -  \mathbb{E}[H_{\alpha}(\hat{\pp})]^2\\
 =& \frac{1}{(1-\alpha)^2} \sum_{i \in \langle V_{\hat{\pp}} \rangle}\sum_{j \in \langle V_{\hat{\pp}} \rangle} 
 \left( \langle \hat{p}_i^{\alpha} \hat{p}_j^{\alpha}\rangle - \langle \hat{p}_i^{\alpha} \rangle \langle \hat{p}_j^{\alpha}\rangle \right)\\
 =&\frac{\alpha^2}{(1-\alpha)^2N}  \sum_{i \in \langle  V_{\hat{\pp}} \rangle}  p_i^{2\alpha-1} 
 - \frac{\alpha^2}{4N^2}  \sum_{i \in \langle  V_{\hat{\pp}} \rangle}  p_i^{2\alpha-2}\\
 =& \frac{\alpha^2}{(1-\alpha)^2} \frac{ V^{(2\alpha)}_{\hat{\pp}}}{N} -  \frac{\alpha^2}{4} \frac{V^{(2\alpha-1)}_{\hat{\pp}}}{N^2}
 \end{alignedat}
\end{equation}
where we used that two different symbols $i \neq j$ are independently drawn, thus $\sum_{i,j} \langle \hat{p}_i^{\alpha} \hat{p}_j^{\alpha} \rangle = \sum_{i \neq j} \langle \hat{p}_i^{\alpha}  \rangle \langle \hat{p}_j^{\alpha} \rangle + \sum_{i} \langle \hat{p}_i^{2\alpha} \rangle$.

\subsection{$D_{\alpha}$}
For $D_{\alpha}$ we have two samples $\hat{\pp}$ and $\hat{\qq}$ each of size $N$ randomly sampled from the distributions $\pp$ and $\qq$ such that we can express the mean and the variance from the expectation values of the corresponding individual entropies.

Introducing the notation $\PP \equiv \frac{1}{2}(\pp+\qq)$ we get for the mean 
\begin{equation}
\begin{alignedat}{2}
  \mathbb{E}[D_{\alpha}(\hat{\pp},\hat{\qq})] =&  \mathbb{E}[H_{\alpha}(\hat{\PP})] - \frac{1}{2}\mathbb{E}\left[ H_{\alpha}\left(\hat{\pp}\right) \right] - \frac{1}{2}\mathbb{E}\left[ H_{\alpha}\left(\hat{\qq}\right) \right]\\
  =&\frac{1}{1-\alpha} \left\{ V_{\hat{\PP}}^{(\alpha+1)} - \frac{1}{2}V_{\hat{\pp}}^{(\alpha+1)} - \frac{1}{2}V_{\hat{\qq}}^{(\alpha+1)}\right\}\\
  &+  \frac{\alpha}{2N} \left\{ \frac{1}{2}V_{\hat{\pp}}^{(\alpha)} +  \frac{1}{2}V_{\hat{\qq}}^{(\alpha)} - \frac{1}{2} V_{\hat{\PP}}^{(\alpha)}\right\}.
  \end{alignedat}
\end{equation}
where $V_{\hat{\PP}}^{(\alpha)}$ denotes the generalized vocabulary, Eq.~\eqref{eq.vocab.alpha}, for the combined sequence $\hat{\PP}=\frac{1}{2}(\hat{\pp}+\hat{\qq})$, which is of length $2N$.

For the variance we get
\begin{equation}
\begin{alignedat}{2}
  \mathbb{V}[D_{\alpha}\left(\hat{\pp},\hat{\qq}\right)] \equiv & \mathbb{E}[D_{\alpha}\left(\hat{\pp},\hat{\qq}\right)^2] - \mathbb{E}[D_{\alpha}\left(\hat{\pp},\hat{\qq}\right)]^2\\
  =& \mathbb{V}[H_{\alpha}(\hat{\PP})] + \frac{1}{4} \mathbb{V}\left[H_{\alpha}(\hat{\pp})\right] + \frac{1}{4}\mathbb{V}\left[H_{\alpha}(\hat{\qq}) \right] \\
  &- \mathrm{Cov}\left[H_{\alpha}(\hat{\PP}), H_{\alpha}(\hat{\pp})+H_{\alpha}(\hat{\qq})  \right],
  \end{alignedat}
\end{equation}
where $ \mathrm{Cov}\left[X,Y\right] \equiv \mathbb{E}[XY] -\mathbb{E}[X]\mathbb{E}[Y]$.
We evaluate the covariance-term in two different ways, i.e.
\begin{equation}
\begin{alignedat}{2}
& (1-\alpha)^2 \mathrm{Cov}\left[H_{\alpha}(\hat{\PP}), H_{\alpha}(\hat{\pp})+H_{\alpha}(\hat{\qq})  \right] \\
=&  \left\langle \sum_{i:\hat{p}_i+\hat{q}_i>0 }  \hat{P}_i^{\alpha}   \left( \sum_{j:\hat{p}_j>0 } \hat{p}_j^{\alpha}+ 	\sum_{j:\hat{q}_j>0 } \hat{q}_j^{\alpha}   \right) \right\rangle  \\
& - \left\langle \sum_{i:\hat{p}_i+\hat{q}_i>0 }  \hat{P}_i^{\alpha} \right\rangle   \left( \left\langle \sum_{j:\hat{p}_j>0 } \hat{p}_j^{\alpha} \right\rangle + \left\langle \sum_{j:\hat{q}_j>0 } \hat{q}_j^{\alpha} \right\rangle  \right) \\
  =&  \left\langle \sum_{i:\hat{p}_i+\hat{q}_i>0 }  \hat{P}_i^{\alpha}   \sum_{j:\hat{p}_j+\hat{q}_j>0 } \left( \hat{p}_j^{\alpha}+ 	 \hat{q}_j^{\alpha}   \right) \right\rangle \\
 &- \left\langle \sum_{i:\hat{p}_i+\hat{q}_i>0 }  \hat{P}_i^{\alpha} \right\rangle    \left\langle \sum_{j:\hat{p}_j+\hat{q}_j>0 } \left( \hat{p}_j^{\alpha} + \hat{q}_j^{\alpha}\right) \right\rangle  \\
  =& \sum_{i \in \left\langle V_{ \hat{\PP} } \right\rangle }\left\{ \left\langle  \hat{P}_i^{\alpha}   \left( \hat{p}_i^{\alpha}+  \hat{q}_i^{\alpha}   \right) \right\rangle - \left\langle \hat{P}_i^{\alpha} \right\rangle   \left( \left\langle \hat{p}_i^{\alpha} \right\rangle + \left\langle \hat{q}_i^{\alpha}  \right\rangle  \right) \right\} 
  \end{alignedat}
\end{equation}
and 
\begin{equation}
\begin{alignedat}{2}
& (1-\alpha)^2 \mathrm{Cov}\left[H_{\alpha}(\hat{\PP}), H_{\alpha}(\hat{\pp})+H_{\alpha}(\hat{\qq})  \right]\\
 =&    \left\langle \sum_{i:\hat{p}_i+\hat{q}_i>0 }  \hat{P}_i^{\alpha} \sum_{j:\hat{p}_j>0 } \hat{p}_j^{\alpha}\right\rangle -  \left\langle \sum_{i:\hat{p}_i+\hat{q}_i>0 }  \hat{P}_i^{\alpha} \right\rangle  \left\langle \sum_{j:\hat{p}_j>0 } \hat{p}_j^{\alpha} \right\rangle \\
  &+    \left\langle \sum_{i:\hat{p}_i+\hat{q}_i>0 }  \hat{P}_i^{\alpha} \sum_{j:\hat{q}_j>0 } \hat{q}_j^{\alpha}\right\rangle -  \left\langle \sum_{i:\hat{p}_i+\hat{q}_i>0 }  \hat{P}_i^{\alpha} \right\rangle  \left\langle \sum_{j:\hat{q}_j>0 } \hat{q}_j^{\alpha} \right\rangle \\
  =&    \sum_{i \in \left\langle V_{ \hat{\pp}} \right\rangle }\left\{ \left\langle \hat{P}_i^{\alpha} \hat{p}_i^{\alpha} \right\rangle 
 - \left\langle \hat{P}_i^{\alpha} \right\rangle  \left\langle \hat{p}_i^{\alpha} \right\rangle \right\}\\
 &+   \sum_{i \in \left\langle V_{ \hat{\qq}} \right\rangle } \left\{ \left\langle \hat{P}_i^{\alpha} \hat{q}_i^{\alpha} \right\rangle 
 - \left\langle \hat{P}_i^{\alpha} \right\rangle  \left\langle \hat{q}_i^{\alpha} \right\rangle \right\}
  \end{alignedat}
\end{equation}
Similarly as in Eq.~\eqref{eq.pialpha.avg} we can approximate
\begin{equation}
\begin{alignedat}{2}
 \left\langle \hat{P}_i^{\alpha}\right\rangle \approx &
  P_i^{\alpha} +\frac{\alpha(\alpha-1)}{4N} P_i^{\alpha-1},\\
\left\langle  \hat{P}_i^{\alpha} \hat{p}_i^{\alpha} \right\rangle
\approx &   P_i^{\alpha}p_i^{\alpha} + \frac{\alpha}{4N}(3\alpha-1)P_i^{\alpha-1}p_i^{\alpha}\\
 &+ \frac{\alpha}{2N}(\alpha-1)P_i^{\alpha}p_i^{\alpha-1},\\
\left\langle  \hat{P}_i^{\alpha} \hat{q}_i^{\alpha} \right\rangle
\approx & P_i^{\alpha}q_i^{\alpha} + \frac{\alpha}{4N}(3\alpha-1)P_i^{\alpha-1}q_i^{\alpha}\\
 &+ \frac{\alpha}{2N}(\alpha-1)P_i^{\alpha}q_i^{\alpha-1}.
  \end{alignedat}
\end{equation}

\begin{widetext}
From this we get for the variance of $D_{\alpha}$
\begin{equation}
\begin{alignedat}{2}
  \mathbb{V}[D_{\alpha}(\hat{\pp},\hat{\qq})] =&\sum_{i \in \langle V_{ \hat{\PP} } \rangle } \left\{ \frac{\alpha^2}{(1-\alpha)^2}\frac{1}{2N}  P_i^{\alpha-1}\left[  P_i^{\alpha} - \frac{1}{2}\left(p_i^{\alpha}+q_i^{\alpha} \right)  \right] - \frac{\alpha^2}{16N^2}  P_i^{\alpha-1}\left[  P_i^{\alpha-1} - \left(p_i^{\alpha-1}+q_i^{\alpha-1} \right)  \right] \right\}\\
  &+ \frac{1}{2} \sum_{i \in \langle V_{ \hat{\pp} } \rangle } \left\{ \frac{\alpha^2}{(1-\alpha)^2}\frac{1}{2N} p_i^{\alpha} \left[ p_i^{\alpha-1} -  P_i^{\alpha-1} \right] - \frac{\alpha^2}{8N^2}p_i^{\alpha-1} \left[ p_i^{\alpha-1} -  P_i^{\alpha-1} \right]   \right\} \\
    &+ \frac{1}{2} \sum_{i \in \langle V_{ \hat{\qq} } \rangle } \left\{ \frac{\alpha^2}{(1-\alpha)^2}\frac{1}{2N} q_i^{\alpha} \left[ q_i^{\alpha-1} -  P_i^{\alpha-1} \right] - \frac{\alpha^2}{8N^2}q_i^{\alpha-1} \left[ q_i^{\alpha-1} -  P_i^{\alpha-1} \right]   \right\}.
  \end{alignedat}
\end{equation}
\end{widetext}

Now we can see that for $\pp=\qq=\PP$  we get
\begin{equation}
 \mathbb{V}[D_{\alpha}(\hat{\pp},\hat{\qq})]_{\pp=\qq} =  \sum_{i \in \langle V_{ \hat{\PP} } \rangle } \frac{1}{16N^2}\alpha^2p_i^{2\alpha-2} = \frac{\alpha^2}{16N^2} V^{(2\alpha-1)}_{\hat{\PP}}.
\end{equation}
While for arbitrary $\pp$ and $\qq$ the variance of the $D_{\alpha}$ contains the variances of the individual entropies (e.g. $V^{(2\alpha)}_{\hat{\PP}}/N $) and a covariance term, (only) in the special case $\pp=\qq$ all first-order terms ($1/N$) vanish yielding a qualitatively different behaviour  $V^{(2\alpha-1)}_{\hat{\PP}}/N^2 $.

\subsection{$\tilde{D}_{\alpha}$}
The finite size estimation of $\tilde{D}_{\alpha}$ can be obtained approximately by
\begin{equation}
\tilde{D}_{\alpha}(\hat{\pp},\hat{\qq})= \frac{D_{\alpha}(\hat{\pp},\hat{\qq})}{D_{\alpha}(\hat{\pp},\hat{\qq})_{\max}} \approx  \frac{D_{\alpha}(\hat{\pp},\hat{\qq})}{\mathbb{E}\left[D^{\max}_{\alpha}(\hat{\pp},\hat{\qq})\right]}
\end{equation}
such that
\begin{equation}
\begin{alignedat}{2}
\mathbb{E}\left[ \tilde{D}_{\alpha}(\hat{\pp},\hat{\qq}) \right] &\approx  \frac{ \mathbb{E} \left[D_{\alpha}(\hat{\pp},\hat{\qq})\right]}{\mathbb{E}\left[D^{\max}_{\alpha}(\hat{\pp},\hat{\qq})\right]},\\
\mathbb{V}\left[ \tilde{D}_{\alpha}(\hat{\pp},\hat{\qq}) \right] &\approx  \frac{ \mathbb{V} \left[D_{\alpha}(\hat{\pp},\hat{\qq})\right]}{\mathbb{E}\left[D^{\max}_{\alpha}(\hat{\pp},\hat{\qq})\right]^2}.
\end{alignedat}
\end{equation}
The mean of $D_{\alpha}^{\max}$ is given according to Eq.~\eqref{eq.Dmax} as a linear combination of the individual entropies of $\hat{\pp}$ and $\hat{\qq}$
\begin{equation}
\begin{alignedat}{2}
 &\mathbb{E}\left[D^{\max}_{\alpha}(\hat{\pp},\hat{\qq})\right] \\
 =&  \dfrac{2^{1-\alpha}-1}{2} \left( \mathbb{E}\left[H_{\alpha} \left( \hat{\pp} \right) \right] + \mathbb{E}\left[H_{\alpha} \left( \hat{\qq} \right) \right]+ \dfrac{2}{1-\alpha}\right).
\end{alignedat}
 \end{equation}

\section{Derivation of Eq.~(\ref{eq.jsd.halpha.bias.2})}
\label{appendixC}

In this section we derive the scaling of the generalized vocabulary $V^{(\alpha)}$ defined in Eq.~\eqref{eq.vocab.alpha} assuming that $\pp$ is a power-law of the form $p_i \propto i^{-\gamma}$, Eq.~(\ref{eq.zipf}).
Instead of looking at the probability of individual symbols $i$, we consider the distribution of frequencies $n$, which in this case yields  $p(n) \propto n^{-1-1/\gamma}$~\cite{newman.2005}.
Consider the sum
$\sum_{i \in V} p_i = \frac{1}{N}\sum_{i \in V} n_i = \frac{1}{N} S_V(\gamma)$, where
$S_V(\gamma)$ corresponds to the sum of $V$ i.i.d. random variables $n_i$ ($i=1,\ldots,V$)
drawn from the distribution $p(n)$
It can be shown that~\cite{bouchaud.1990}
\begin{equation}
 S_V(\gamma) \propto \begin{cases}
                      V^{\gamma}, & \gamma>1,\\
                      V, & \gamma<1.
                     \end{cases}
\end{equation}
The case $\gamma=1$ includes additional logarithmic corrections, but is not of relevance for the discussion, therefore, we leave it for sake of simplicity.
In the same way, we can treat $\sum_{i \in V} p_i^{\mu} = \frac{1}{N^{\mu}}\sum_{i \in V}
n_i^{\mu} = \frac{1}{N^{\mu}} S_V(\gamma \mu)$ by noting that $S_V(\gamma \mu)$ can be
interpreted as the sum of $V$ i.i.d. random variables $n_i$ ($i=1,\ldots,V$), where $n_i
\sim \tilde{p}(n)$ with $\tilde{p}(n) \propto n^{-1-1/(\gamma \mu)}$ such that we get 
\begin{equation}
 S_V(\gamma \mu) \propto \begin{cases}
                      V^{\gamma \mu}, & \mu<1/\gamma,\\
                      V, & \mu>1/\gamma.
                     \end{cases}
\end{equation}
By setting $\mu =\alpha-1$ in Eq.~(\ref{eq.vocab.alpha}) and noting that for $p_i\propto i^{-\gamma}$, Eq.~(\ref{eq.zipf}), the number of different symbols scales
as $V\propto N^{1/\gamma}$, Eq.~(\ref{eq.heaps}), we obtain Eq.~(\ref{eq.jsd.halpha.bias.2}).

\section{Temporal evolution of $\tilde{D}_\alpha(\Delta t)$}
\label{appendixD}
We are interested in understanding the dependence of $\tilde{D}_\alpha(\Delta t)\equiv
\tilde{D}_\alpha(t_0, t_0+\Delta t)$ on $\Delta t$ (we assume $\tilde{D}_\alpha(\Delta t)$
is the same for all $t_0$).  The triangle inequality implies that
\begin{equation}\label{eq.bound}
\begin{alignedat}{2}
\tilde{D}_\alpha(\Delta t) \le & \left( \sqrt{\tilde{D}_\alpha(\Delta t-1)} 
+\sqrt{\tilde{D}_\alpha(\Delta t=1)} \right)^2 
\\ \le & (\Delta t)^2\tilde{D}_\alpha(\Delta t=1).
\end{alignedat}
\end{equation}
In order to consider the origin of different $\Delta t$ dependencies within the general
bound given by Eq.~(\ref{eq.bound}), we consider two classes of words subject to frequency
change in $\Delta t$: (i)  words which show fluctuations (e.g., finite sampling or topical dependencies) that do not depend on $\Delta t$; and (ii) words which show a systematic increase or decrease over all $t$.
If we assume that all words that change fall in one of these classes we can use the fact that $\tilde{D}_\alpha$ is defined as a sum of word
types and decompose the total change $\tilde{D}_\alpha$ as  $\tilde{D}_\alpha = \tilde{D}^{(i)}_\alpha +
\tilde{D}^{(ii)}_\alpha$, where $\tilde{D}^{(i,ii)}_\alpha$ is the divergence of all words
falling in class (i) or (ii), respectively. 
For category (ii), the changes between consecutive years are independent, thus, the equality case of the
triangle inequality  is obtained: $\sqrt{\tilde{D}_\alpha(t_0, t_2)} = \sqrt{\tilde{D}_\alpha(t_0,
  t_1)}+\sqrt{\tilde{D}_\alpha(t_1, t_2)}$ for all $t_1$ with $t_0 \le t_1 \le t_2$.
Therefore we obtain a quadratic dependence on $\Delta t$ as 
\begin{equation}\label{eq.linear}
\tilde{D}_\alpha(\Delta t) = \tilde{D}^{(i)}_\alpha + \tilde{D}^{(ii)}_\alpha(\Delta t=1) (\Delta t)^2 
\end{equation}

%%%%%%%%%%%%%%%%%%%%%%%%%%%%%%%%% Appendix %%%%%%%%%%%%%%%%%%%%%%%%%%%%%%%%%
% \bibliographystyle{longbibliography}
\bibliography{bib_jsd}

\end{document}